\documentclass[12pt]{article}
\usepackage{amsmath, amsfonts, amscd}

\newcommand{\be}[1]{\begin{equation}\label{#1}}
\newcommand{\bra}[1]{{\langle #1 |}}
\newcommand{\braket}[2]{{\langle #1 |\, #2\rangle}}

\newcommand{\ddd}{{D}}

\newcommand{\ee}{\end{equation}}

\newcommand{\ket}[1]{{|\, #1\rangle}}
\newcommand{\ketbra}[2]{{|\, #1\rangle\langle #2|}}

\newcommand{\omg}{\Omega}

\newcommand{\spn}{{\rm span}}

\newcommand{\amg}{\vec{\Omega}}
\newcommand{\ess}{\vec{S}}
\newcommand{\imm}{\stackrel{\ast}{\rightarrow}}

\newcommand{\pot}{\vec{\phi}_{n}}
\newcommand{\impot}{\pot^{\ast}}

\newcommand{\N}{\mathbb{N}}
\newcommand{\real}{\mathbb{R}}
\newcommand{\com}{\mathbb{C}}
\newcommand{\quat}{\mathbb{H}}
\newcommand{\integer}{\mathbb{Z}}

\newcommand{\cem}{{\cal{M}}}

\title{\bf\Large Finitary Spacetime Sheaves of Quantum\\ Causal Sets:
Curving Quantum Causality}

\author{A. Mallios\thanks{Algebra and Geometry Section,
Department of Mathematics, University of Athens,
Panepistimioupolis 157 84, Athens, Greece; e-mail:
amallios@cc.uoa.gr} and I. Raptis\thanks{Department of
Mathematics, University of Pretoria, Pretoria 0002, Republic of
South Africa; e-mail: iraptis@math.up.ac.za}}

\date{}

\begin{document}

\maketitle

\begin{abstract}

\noindent A locally finite, causal and quantal substitute for a
locally Minkowskian principal fiber bundle $\cal{P}$ of modules of
Cartan differential forms $\omg$ over a bounded region $X$ of a
curved $C^{\infty}$-smooth differential manifold spacetime $M$
with structure group ${\bf G}$ that of orthochronous Lorentz
transformations $L^{+}:=SO(1,3)^{\uparrow}$, is presented.
${\cal{P}}$ is the structure on which classical Lorentzian
gravity, regarded as a Yang-Mills type of gauge theory of a
$sl(2,\com)$-valued connection $1$-form $\cal{A}$, is usually
formulated. The mathematical structure employed to model this
replacement of ${\cal{P}}$ is a principal finitary spacetime
sheaf $\vec{\cal{P}}_{n}$ of quantum causal sets $\amg_{n}$ with
structure group ${\bf G}_{n}$, which is a finitary version of the
group  ${\bf G}$ of local symmetries of General Relativity, and a
finitary Lie algebra ${\bf g}_{n}$-valued connection $1$-form
${\cal{A}}_{n}$ on it, which is a section of its sub-sheaf
$\amg^{1}_{n}$. ${\cal{A}}_{n}$ is physically interpreted as the
dynamical field of a locally finite quantum causality, while its
associated curvature ${\cal{F}}_{n}$, as some sort of `finitary
Lorentzian quantum gravity.

\end{abstract}

\newpage
\begin{quotation}
{\footnotesize ...The locality principle seems to catch something fundamental
about nature... Having learned that the world need not be Euclidean in the large,
the next tenable position is that it must at least be Euclidean in the small,
a manifold. The idea of infinitesimal locality presupposes that the world is
a manifold. But the infinities of the manifold (the number of events per unit
volume, for example) give rise to the terrible infinities of classical field
theory and to the weaker but still pestilential ones of quantum field theory.
The manifold postulate freezes local topological degrees of freedom which are
numerous enough to account for all the degrees of freedom we actually observe.

The next bridgehead is a dynamical topology, in which even the local
topological structure is not constant but variable. The problem of
enumerating all topologies of infinitely many points is so absurdly
unmanageable and unphysical that dynamical topology virtually forces us to a
more atomistic conception of causality and space-time than the continuous
manifold... (D. Finkelstein, 1991)}
\end{quotation}

\section*{\normalsize\bf 1. INTRODUCTION CUM PHYSICAL MOTIVATION}

\indent\normalsize\sloppy We are still in need of a cogent quantum theory
of gravity. A quantum field theoretic scenario for General Relativity (GR)
is assailed by non-renormalizable infinities coming from the singular
values of fields
that are assumed to propagate and interact on a smooth spacetime manifold.
Most likely, it is
our modeling of spacetime after a $C^{\infty}$-smooth differential
manifold that is the culprit for this unpleasant situation. We can hardly
expect Nature to have any infinities, but we can be almost certain that it is our own
theoretical models of Her that are of limited applicability and validity.

The present paper takes a first step towards arriving at an operationally sound, locally finite, causal
and quantal model of classical Lorentzian gravity from a finitary spacetime
sheaf (finsheaf) theoretic point of view. Classical Lorentzian gravity is regarded as a
Yang-Mills type of gauge theory of a $sl(2,\com)$-valued connection
$1$-form ${\cal{A}}$ that is suitably formulated on a locally Minkowskian principal fiber
bundle $\cal{P}$ of modules of Cartan differential forms $\omg$ over a bounded
region $X$ of a curved $C^{\infty}$-smooth differential manifold spacetime $M$ with
structure group ${\bf G}$ that of orthochronous Lorentz transformations
$L^{+}:=SO(1,3)^{\uparrow}$.
A principal finsheaf $\vec{\cal{P}}_{n}$ of quantum causal
sets (qausets\footnote{Since `causal sets' are coined `causets' for short
by Sorkin (private communication), `quantum causal sets' may be similarly
nicknamed `qausets'.}) $\amg_{n}$ having as structure group a finitary version
${\bf G}_{n}$ of $L^{+}$, together with a finitary spin-Lorentzian connection
${\cal{A}}_{n}$ which is a ${\bf g}_{n}$-valued section of the sub-sheaf $\amg^{1}_{n}$
of reticular $1$-forms of $\vec{\cal{P}}_{n}$,
is suggested as a locally finite model, of strong operational character, of the
dynamics of the quantum causal relations between events and their local causal
symmetries in a bounded region $X$ of a curved smooth spacetime manifold $M$.
In short, we propose
$(\vec{\cal{P}}_{n},{\cal{A}}_{n})$ as a finitary, causal and quantal replacement
of the classical gravitational spacetime structure $({\cal{P}},{\cal{A}})$\footnote{Our scheme
may be loosely coined a `finitary Lorentzian quantum gravity', although it is more precise
to think of $\vec{\cal{P}}_{n}$ as a finitary, causal and quantal substitute for the
structure $\cal{P}$ on which GR is cast as a gauge theory,
rather than directly of GR on it {\it per se}. For instance, we will go as far as
to define curvature ${\cal{F}}_{n}$ on $\vec{\cal{P}}_{n}$, but we will not give an
explicit expression of
the Einstein equations on it. The latter is postponed to another paper (Raptis, 2000{\it f}).}.
The theoretical model $(\vec{\cal{P}}_{n},{\cal{A}}_{n})$ is supposed to be a preliminary
step in yet another attempt at viewing the problem of
`quantum gravity' as the dynamics of a local, finitistic and quantal version
of a variable causality\footnote{See opening quotation above.} (Finkelstein,
1988, 1989, 1991, 1996, Bombelli {\it et al.}, 1987, Sorkin, 1990, 1995, Raptis, 2000{\it f}).

In more detail, the continuous ({\it ie}, $C^{0}$) topology of a bounded region $X$ of a
spacetime manifold $M$ has been successfully approximated by so-called
`finitary topological spaces' which are mathematically modeled after
partially ordered sets (posets) (Sorkin, 1991). The success of such
coarse approximations of the topological spacetime continuum rests on the fact that an
inverse system consisting of finer-and-finer finitary posets possesses, at
the maximum
(finest) resolution of $X$ into its point-events, a limit space that is
effectively homeomorphic to $X$ (Sorkin, 1991).

In a similar way, coarse approximations of the continuous ({\it ie}, $C^{0}$)
spacetime observables on $X$ have been soundly modeled after so-called
`finitary spacetime sheaves' (finsheaves) which are structures consisting of
continuous functions
on $X$ that are locally homeomorphic to the finitary posets of Sorkin
(Raptis, 2000{\it a}). Also, an inverse system of such finsheaves was seen
to `converge', again at maximum refinement and localization of $X$ to
its point-events, to $S(X)$-the sheaf of (germs of sections of) continuous
spacetime observables on $X$ (Raptis, 2000{\it a}).

In (Raptis and Zapatrin, 2000), an algebraic quantization procedure of
Sorkin's finitary poset substitutes for continuous spacetime topology was
presented, first by associating with every such poset $P$ a non-commutative
Rota incidence
algebra $\omg(P)$, then by quantally interpreting the latter's structure. The
aforementioned limit of a net of such quantal incidence algebras was
interpreted as Bohr's correspondence principle in the sense that the
continuous spacetime topology emerges, as a classical structure, from some
sort of decoherence of the underlying discrete and coherently superposing
quantum Rota-algebraic topological substrata (Raptis and Zapatrin, 2000).
The operationally
pragmatic significance of the latter, in contradistinction to the ideal and,
because of it, pathological\footnote{Due to the unphysical infinities in
the form of singularities from which the classical and quantum field
theories, which are defined on the operationally ideal spacetime continuum,
suffer (see also opening quotation).}
event structure that the classical differential manifold model of spacetime
stands for, was also emphasized by Raptis and Zapatrin.

Furthermore, it has been argued (Raptis and Zapatrin, 2000) that,
in view of the fact that the $\omg(P)$s were seen to be
discrete differential manifolds in the sense of Dimakis and M\"uller-Hoissen
(1999), not only the continuous ($C^{0}$) topological, but also the smooth
({\it ie}, $C^{\infty}$) differential structure of classical spacetime,
emerges at the operationally ideal classical limit of finest resolution of
a net of quantal incidence algebras. Since only at this ideal classical limit
of an inverse system of such quantum topological substrata the local
structure of the differential spacetime manifold emerges\footnote{That is to
say, the spacetime event and the space of covariant directions tangent to
it ({\it ie}, its cotangent space of differential forms).}, the substrata
were conceived
as being essentially alocal structures (Raptis and Zapatrin, 2000), with
this `a-locality' signifying some sort of independence of these algebraic
structures from the classical conception of spacetime as a smooth background
geometric base space. Similarly, the finsheaf theoretic approach developed
in (Raptis, 2000{\it a}), with its finitary algebraic-operational character,
strongly emphasizes the physical significance of
such a non-commitment to an inert background geometrical base spacetime
manifold, as well as its accordance with the general operational, ultimately pragmatic,
philosophy of quantum theory (Finkelstein, 1996).

Moreover, at the end of (Raptis, 2000{\it a}), it
is explicitly mentioned that by assuming further algebraic structure for the
stalks of the aforementioned finsheaves, as for instance by considering
sheaves of incidence algebras over Sorkin's finitary topological posets,
at the limit of maximum resolution of a net
of such finsheaves of Rota algebras, which can also be regarded as Bohr's
classical limit {\it \`{a} la} Raptis and Zapatrin (2000), the differential triad
$(X,\omg :=\oplus_{i}\omg^{i},\ddd)$ should emerge. The latter
stands for the sheaf of modules
of Cartan differential forms $\omg$ on the smooth $X$, equipped with the
K\"ahler-Cartan differential operator $\ddd$ which effects (sub)sheaf morphisms
of the following sort $\ddd :~ (X,\omg^{i})\rightarrow(X,\omg^{i+1})$
(Mallios, 1998). Thus, a finsheaf of
Rota incidence algebras is expected to be a sound model of locally finite,
as well as quantal, `approximations' of the smooth spacetime observables-the
classical spacetime dynamical fields\footnote{We tacitly assume that the classical model for
spacetime and the fields inhabiting, propagating and interacting on it is that of a $4$-dimensional
differential (or $C^{\infty}$-smooth) manifold, with fiber-spaces $\omg^{n}$
of smooth Cartan exterior $n$-forms attached (in fact, $0\leq n\leq 4$, but
we will not be further concerned about questions of dimensionality; see discussion in
$(c^{'})$ of section 6). Physical fields are then
modeled after cross-sections of this Cartan fiber bundle ${\cal{P}}$
of smooth exterior
forms (G\"ockeler and Sch\"ucker, 1990, Baez and Muniain, 1994).}.

Parenthetically, and with an eye towards the physical
interpretation to be given subsequently to our mathematical model, we should
mention at this point that the inverted commas
over the word `approximations' in the last sentence above may be explained
as follows: after the successful algebraic quantization
of Sorkin's discretized spacetimes in (Raptis and Zapatrin, 2000),
it has become clear that the resulting alocal quantum
topological incidence algebras $\omg(P)$ associated with the finitary
topological posets $P$ in (Sorkin, 1991) should not be thought of as
approximations-proper of the classical smooth differential forms like
their corresponding $P$s or the finsheaves $S_{n}$ in
(Raptis, 2000{\it a}) approximate the $C^{0}$-topological manifold structure
of classical spacetime, as if a geometric spacetime exists as a background
base space `out there'.
Rather, they should be regarded as operationally pragmatic and relatively autonomous
quantum spacetime structures an inverse system\footnote{`Collection' or even
`ensemble' are also appropriate synonyms to `system' in this context.}
of which possesses an operationally ideal ({\it ie},
unobservable in actual experiments) and classical, in the sense of Bohr,
limit structure isomorphic
to the differential manifold model of spacetime (Raptis and Zapatrin, 2000).
>From this viewpoint, the quantum
topological incidence algebras $\omg(P)$ (and their qauset relatives in
(Raptis, 2000{\it b})) are regarded as being physically fundamental
(primary) and their correspondence limit geometric manifold structure
as being derivative
(secondary), ultimately, their emergent classical counterpart in much the
same way that the classical Poisson algebra of observables on the geometric
phase (state) space of a classical mechanical system is the `classical decoherence
limit'\footnote{That is to say,
when coherent quantum superpositions between observables are lifted.} or the
`classical undeformation'\footnote{That is to say, the undoing of the
formal procedure of `quantum deformation'.} of the
Heisenberg algebra of an underlying quantum system in the operationally
ideal situation of non-interfering quantally and non-perturbing dynamically\footnote{That is to say, infinitely smooth.} operations of
observation (determination) of the properties of
the latter\footnote{Altogether, formally `as $\hbar\rightarrow 0$'.}.
Properly conceived,
it is the classical theory (model) that should be thought of as an
approximation of the deeper quantum theory (model), not the other way
around (Finkelstein, 1996). Thus, `quantum replacements' or `quantum
substitutes' instead of `approximations' will be used more often from now on
to describe our finsheaves (of qausets), although they were initially conceived as
approximations-proper of the continuous spacetime topology in (Raptis,
2000{\it a}) as it was originally motivated by Sorkin (1991). In total,
this non-acceptance of ours of spacetime as an inactive smooth geometric
receptacle of the physical fields or as a background stage that supports their
dynamical propagation, that is passively existing as a static state space
`out there' and whose structure is fixed {\it a priori} independently of our
experimental actions on or operations of observation of `it', is the essence
of the operationally sound quantum physical
semantics that we will give to our algebraic finsheaf model in the present
paper.

In GR, the classical theory of gravity which is based on the structural
assumption that spacetime is a $4$-dimensional pseudo-Riemannian manifold $M$,
the main dynamical variable is the smooth Lorentzian spacetime metric $g_{\mu\nu}$
which is physically interpreted as the gravitational potential. The local
relativity group of GR, in its original formulation in terms of the Lorentzian
metric $g_{\mu\nu}$, is the orthochronous Lorentz group
$L^{+}:=SO(1,3)^{\uparrow}$. GR may also be
formulated in terms of differential forms on the locally Minkowskian bundle
${\cal{P}}$ (G\"ockeler and Sch\"ucker, 1990)\footnote{See chapter on Einstein-Cartan
theory. We call ${\cal{P}}$ `the Cartan principal fiber bundle with structure
group the orthochronous Lorentz group $L^{+}$ of local
invariances of GR'. See sections 2 and 5.}. Equivalently, in its gauge
theoretic spinorial formulation
(Bergmann, 1957\footnote{In this theory, $g_{\mu\nu}$ is replaced
by a field of four $2\times 2$ Pauli spin-matrices.},
Baez and Muniain, 1994\footnote{We refer to Ashtekar's modification of the
Palatini formulation of GR by using new spin
variables (Ashtekar, 1986). In this theory, only the self-dual part
${\cal{A}}^{+}$ of a spin-Lorentzian connection ${\cal{A}}$ is regarded as
being physically significant. In (Raptis, 2000{\it e}) this is used as an example to argue that the fundamental quantum time asymmetry expected of `the true quantum gravity' (Penrose, 1987) is already built into the kinematical structure of a locally finite, causal and quantal version of that theory modeled after curved finitary spacetime sheaves (or schemes) of qausets.} type of gauge theory
of a $sl(2,\com)$-valued $1$-form ${\cal{A}}$-the spin-Lorentzian connection
field, which represents the gravitational gauge potential. A sound model
for this theory is a principal fiber bundle ${\cal{P}}$ over (the region $X$ of)
the $C^{\infty}$-smooth spacetime manifold $M$,
with structure group ${\bf G}=SL(2,\com)$\footnote{A principal fiber
bundle with structure group ${\bf G}$ may also be called a
`${\bf G}$-bundle' for short.} and a non-flat connection $1$-form
${\cal{A}}$ taking values in the Lie algebra ${\bf g}=sl(2,\com)$ of
${\bf G}$\footnote{Since locally in the group fiber ({\it ie}, Lie
algebra-wise in the fiber space) of the ${\bf G}$-bundle ${\cal{P}}$
$sl(2,\com)$ is isomorphic to the Lie algebra $\ell^{+}=so(1,3)^{\uparrow}$
of the orthochronous Lorentz group $L^{+}$, ${\cal{P}}$
may equivalently be thought of as having the latter as structure group
${\bf G}$. Due to this local isomorphism, ${\cal{A}}$ is given the epithet
`spin-Lorentzian' and the same symbol ${\cal{P}}$ is
used above for both the Cartan (Lorentzian) and the Bergmann (spin) ${\bf G}$-bundles.
Thus, ${\cal{P}}$ is called `the Cartan-Bergmann ${\bf G}$-bundle'. See
section 5 for more on this local isomorphism between the Cartan and the
Bergmann ${\bf G}$-bundles.}, totally,
$(X,{\cal{P}},{\bf G},{\cal{A}})$\footnote{The name `principal' is usually
reserved only for the group ${\bf G}$-bundle or sheaf, while the vector or algebra sheaf that carries it, in
our case $\omg$, is called `associated' (Mallios, 1998). Here we use one symbol, ${\cal{P}}$, and one
name, `principal', for both the ${\bf G}$-sheaf of orthochronous Lorentz transformations $L^{+}$
and its associated locally Minkowskian sheaf of differential forms $\omg$. Conversely, in section 4 we
first define $\omg$ as an algebra sheaf and then we coin the ${\bf G}$-sheaf of its Lorentz
symmetries `adjoint'. There is no misunderstanding: $\omg$ is associated with ${\bf G}$, or vice versa,
${\bf G}$ is adjoint to $\omg$, and together they constitute the principal sheaf ${\cal{P}}$. Nevertheless, we
apologize to the mathematical purist for this slight change in nomenclature.}. Thus, by the discussion in the
penultimate paragraph, it follows that a principal finsheaf of
quantum incidence algebras, together with a non-flat connection taking values in
their local symmetries, may be employed to model a locally finite and
quantal version of Lorentzian gravity in its gauge theoretic formulation on
a smooth spacetime manifold.

However, there seem to be {\it a priori} two serious problems with such
a model. On the one
hand, only Riemannian ({\it ie}, positive definite) metric connections may
be `naturally' defined on discrete differential manifolds such as our Rota
incidence algebras (Dimakis and M\"uller-Hoissen, 1999), and on the other,
the anticipated classical limit sheaf or fiber bundle $(X,\omg ,\ddd)$
is flat (Mallios,
1998)\footnote{Dimakis and M\"uller-Hoissen (1999) also mention the fact that
the (torsionless) Riemannian metric connection $\nabla$ of the universal differential
calculus on a discrete differential manifold is flat, in that it reduces
to the nilpotent K\"ahler-Cartan differential $\ddd$ whose curvature ${\cal{R}}$ is zero,
since ${\cal{R}}:=\nabla^{2}=\ddd^{2}=0$.}. The first comes into conflict
with the indefinite character of the local spacetime metric of GR\footnote{In
GR, the local metric field
$g_{\mu\nu}$ is Lorentzian (of signature $2$), not Euclidean (of trace $4$).},
thus also with its local relativity group\footnote{The group of local
isometries of GR, at least in its spinorial gauge theoretic
formulation mentioned
above, is taken to be $SL(2,\com)$-the double cover of the orthochronous
Lorentz group $L^{+}=SO(1,3)^{\uparrow}$ of local invariances of
$g_{\mu\nu}$ that also locally preserve the orientation of time, not the
$4$-dimensional unimodular Euclidean
rotations in $SO(4)$. In this sense GR is a theory of
(locally) Lorentzian gravity (see next section and 5).}; while the second, with the general relativistic
conception of the gravitational field strength as the non-vanishing
curvature of spacetime.

One should not be discouraged, for there seems to be a way out of this
double impasse which essentially motivated us to consider finsheaves of qausets
in the first place. First, to deal with the `signature problem',
we must change physical interpretation for the algebraic structure of the
stalks of the aforementioned
finsheaf of quantal incidence algebras from `topological' to `causal'. This
means that we should consider finsheaves of the qausets
in (Raptis, 2000{\it b}), rather than finsheaves of the quantum discretized
spacetime topologies in (Raptis and Zapatrin, 2000). Indeed, Sorkin (1995),
in the context of constructing a plausible theoretical model for quantum gravity,
convincingly argues for a physical interpretation of finitary posets as
locally finite causets (Bombelli {\it et al.}, 1987, Sorkin, 1990) and against their
interpretation as finite topological spaces or simplicial complexes
(Alexandrov, 1956, 1961, Raptis and Zapatrin, 2000). Similar arguments
against a non-relativistic, spatial conception of
topology and for a temporal or causal one which is also algebraically
modeled with a quantum interpretation given to this algebraic structure,
like the quantum causal topology of the qausets in (Raptis, 2000{\it b}),
are presented in (Finkelstein, 1988).
Ancestors of the causet idea are the classic works of Robb (1914),
Alexandrov (1956, 1967) and Zeeman (1964, 1967) which show that the entire
topology and conformal geometry of Minkowski space $\cem$, as well as its
relativity group $L^{+}$ of global orthochronous Lorentz transformations modulo
spacetime volume-preserving maps, can be determined
by modeling the causal relation between its events after a partial order.
Alternatively, the spirit of deriving the entire geometry of the Lorentzian
spacetime manifold from causality modeled after a partial order, is captured
by the following words taken from (Bombelli {\it et al.}, 1987)\footnote{See
also (Bombelli and Meyer, 1989, Sorkin, 1990).}:

\begin{quotation}
{\footnotesize There is a fact, insufficiently appreciated in our view,
that a classical space-time's causal structure comes very close to
determining its entire geometry. By the causal structure of a space-time,
one means the relation $P$  specifying which events lie to the future of
which other events. Ordinarily, one thinks of space-time as a topological
manifold $M$, endowed with a differentiable structure $S$, with respect to
which a metric $g_{ab}$ is defined. Then the causal order $P$ is regarded as
derived from the lightcones of $g$. However, one can also go the other way:
Given a space-time obeying suitable smoothness and causality conditions (and
of dimensionality $>2$), let us retain from all its structure the information
embodied in the order relation $P$. Then we can recover from $P$ not only
the topology of $M$, but also its differentiable structure, and the conformal
metric, $g_{ab}/|det(g)|^{1/n}$. Now a partial ordering is a very simple
thing, and it is natural to guess that in reality $g_{ab}$ should be derived
from $P$ rather than the other way around...}
\end{quotation}

On the other hand, causality as a partial order, while it solves the `signature
problem', is unable to adequately address the second `curvature problem'
mentioned above, since it determines the Minkowski space $\cem$ of Special
Relativity (and its Lorentz symmetries) which is flat (and its Lorentz
symmetries are global). Our way out
of this second `curvature impasse' involves a rather straightforward
localization or gauging of the qausets in (Raptis, 2000{\it b}),
by considering a non-flat connection on a
finsheaf of such quantally and causally interpreted
incidence algebras, thus by emulating the
work of Mallios (1998)\footnote{Albeit, in a finitary causal and quantal context.}
that studies Yang-Mills gauge connections on ${\bf G}$-sheaves of vector
spaces and algebras in general. This gauging of quantum causality translates in
a finitary and quantal setting the fact that the classical theory of gravity,
GR, may be regarded as Special Relativity (SR)-localized or being
gauged\footnote{So that the spacetime metric, or its associated connection,
become dynamical field variables (Torretti, 1981).}. This
connection variable is supposed to represent the
dynamics of an atomistic local quantum causality as the latter is algebraically encoded
stalk-wise in the finsheaf ({\it ie}, in the qausets). The result may be regarded as the first essential step towards formulating a finitary dynamical scenario for the qauset stalks of the sheaf which, in turn, may be
physically interpreted as a finitary and causal
model of the still incompletely or even well-formulated Lorentzian quantum
gravity. Equivalently, and in view of the sound operational interpretation
given to the topological incidence algebras in (Raptis and Zapatrin, 2000)
as well as to the topological finsheaves in (Raptis, 2000{\it a}), our model
may be physically interpreted
as locally finite and quantal replacements of the dynamics of
the local causal relations between events and their local causal
symmetries\footnote{That is to say, the dynamics of local quantum causality
or `local quantum causal topology' and its symmetries.}, in a limited
(finite or bounded), by our own domain of
experimental activity ({\it ie}, laboratory)
(Raptis and Zapatrin, 2000), region $X$ of the smooth spacetime manifold
$M$\footnote{See sections 5 and 6.}.
As we mentioned above, the latter `exists' only as a `surrogate background
space' that helps one remember where the discreteness of our model comes from, but it
is not essential to the physical problem in focus. The spacetime continuum,
as a `base space', is only a geometrical scaffolding that supports our
structures\footnote{In the sense that `it avails itself to us as a topological space'
by providing sufficient (but not necessary !)
conditions for the definition of ${\cal{A}}_{n}$ which is the main dynamical variable
in our theoretical scheme. See section 5.}, but that should also be
discarded after their essentially alocal-algebraic, quantal-operational and
causal ({\it ie}, non-spatial, but temporal) nature is
explicated and used for our problem in focus. Then, the
aforementioned correspondence principle for quantal topological
incidence algebras may be used on (an inverse system of) the principal
finsheaves of qausets and their non-flat spin-Lorentzian connections in
order to recover the classical spacetime structure on which GR is formulated,
as the classical theory of gravity,
at the classical and operationally ideal limit
of resolution ({\it ie}, of infinite localization and infinitesimal/differential
separation) of spacetime into its events. This classical limit spacetime model for
GR, as a gauge theory, is the one mentioned above, namely,
a principal fiber bundle $\cal{P}$ of modules of smooth Cartan differential
forms $\omg$, over (a region $X$ of) a $C^{\infty}$-smooth
Lorentzian spacetime manifold $M$, with structure group ${\bf G}=SL(2,\com)$ or its
locally isomorphic $SO(1,3)^{\uparrow}$, and a non-flat $sl(2,\com)$-valued gravitational
gauge connection $1$-form ${\cal{A}}$ on it\footnote{Thus, ${\cal{A}}$ is a
cross-section of the $\omg^{1}$ sub-bundle of the Cartan-Bergmann ${\bf G}$-bundle
${\cal{P}}$ taking values in $sl(2,\com)\simeq\ell^{+}$.}.

The present paper is organized as follows: in the next section we propose and
discuss in detail finitary versions of the principles of Equivalence and
Locality of GR, as well as of their `corollaries', the
principles of
Local Relativity and Local Superposition, that are expected to be operative at the locally
finite setting that we place our first step at modeling `finitary Lorentzian quantum
gravity' after `curving quantum causality by gauging a principal finsheaf of
qausets'\footnote{The word `gauging' pertains to the aforementioned
implementation of a non-flat gauge connection ${\cal{A}}_{n}$ on the finsheaf
in focus.}. In section 3 we review the algebraic
model of flat quantum causality proposed in (Raptis, 2000{\it b}), namely,
the qauset, and emphasize its local aspects to be subsequently
gauged (in section 5). In section 4 we recall the topological
finsheaves from (Raptis, 2000{\it a}), then we define
finsheaves of qausets and their local symmetries. At the end of
the section, a sound operational interpretation of finsheaves of
qausets and their symmetries is given, so that our theory is
shown to have a strong philosophical support as well. In section 5 we suggest
that for localizing or gauging and, as a result, curving quantum
causality, a principal finsheaf of qausets having as structure group a
finitary version of $SO(1,3)^{\uparrow}$, together with a discrete and local sort
of a non-flat, spin-Lorentzian connection ${\cal{A}}_{n}$ on it,
is an operationally sound model. ${\cal{A}}_{n}$ is then physically
interpreted as a finitary, local, causal and quantal topological variable whose
non-zero curvature stands for a finitary, causal and quantal model of
Lorentzian gravity. We conclude the paper by discussing the soundness of
this finsheaf model of finitary and causal Lorentzian quantum gravity as well as six
physico-mathematical issues that derive from it.

\section*{\normalsize\bf 2. PHYSICAL PRINCIPLES FOR FINITARY LORENTZIAN
QUANTUM GRAVITY}

\indent\sloppy In this section we commence our endeavor to model connection
(and its associated curvature) in a curved finitary quantum causal setting by
establishing heuristic physical principles that must be encoded in the very
structure of our mathematical model\footnote{The `principal finsheaf of
qausets with a non-flat finitary spin-Lorentzian connection on it', to be
built progressively in the next three sections.} on which the dynamics of a
locally finite quantum causality is going to be founded in section 5. The four
physical principles to be suggested
here will be seen to be the finitary and (quantum) causal analogues of the
ones of Equivalence, Locality, as well as their `corollary' principles of
Local Relativity and Local Superposition respectively, of GR
which is formulated as a gauge theory in ${\cal{P}}$ over a differential
manifold spacetime $M$. We have chosen these principles from the theory of
classical gravity, because they show precisely in what way the latter is a
type of gauge theory, and also because they will prepare the reader for our
localization or gauging and curving of qausets in section 5.

The first physical
principle from GR that we would like to adopt
in our inherently reticular scenario, so that curvature may be naturally
implemented and straightforwardly interpreted as gravity in a finitary
(quantum) causal context like ours, is that of
equivalence (EP). We borrow from GR the following
intuitively clear version of the EP:

\noindent\underline{\it Classical Equivalence Principle (CEP)}: the curved spacetime
of GR is locally Minkowskian; thence flat. That is to say, the
space tangent to every spacetime event is isomorphic to flat Minkowski space
$\cem$. As we mentioned in the introduction, in this
sense GR may be thought of as SR made
local or been point-wise (event-wise) gauged. Expressed thus, the CEP
effectively encodes Einstein's fundamental insight that locally the
gravitational field $g_{\mu\nu}$ can be `gauged away' or be reduced to the
constant and flat Minkowski metric $\eta_{\mu\nu}=diag(-1,+1,+1,+1)$ of
SR\footnote{Since $\eta_{\mu\nu}(x)$ delimits the Minkowski
lightcone at $x$ for every $x\in M$, which, in turn, defines the local
causal relations between events in the Minkowski space tangent to $x$,
the gravitational potential $g_{\mu\nu}$ may be alternatively interpreted as
`the dynamical field of local causality'.} by passing to a locally inertial
frame\footnote{Equivalently put, CEP states that a body gravitated under
a constant gravitational field intensity $\Gamma$ is physically
indistinguishable from
a uniformly accelerated one with constant acceleration $\gamma=\Gamma$,
a statement that entails the local equivalence between the body's gravitational
and inertial mass.} (Torretti, 1981).

What is important to emphasize in this
formulation of the CEP is that in GR $\cem$ assumes a
local kinematical role, in the sense that an isomorphic copy of it is
erected, as some kind of `fiber space', over each event of the differential manifold
spacetime, so that every individual fiber is physically interpreted
as `an independent (of the other $\cem$-fibers) vertical world of spacetime
possibilities along which the
dynamically variable $g_{\mu\nu}$ can be reduced to the constant
$\eta_{\mu\nu}$'\footnote{The epithet `kinematical' for CEP may be also
justified as follows: for every point-event $x$ of the
pseudo-Riemannian spacetime manifold $M$ of GR, there is
a coordinate system among all the possible general coordinate frames, a
so-called
locally inertial one (the epithets `normal' or `geodesic' may also be used
instead of `inertial'), with respect to which $g_{\mu\nu}=\eta_{\mu\nu}$ and $(\frac{\partial
g_{\mu\nu}}{\partial x_{\lambda}})=0$, but exactly due to gravity, the second
partial derivatives of the metric cannot be made to vanish and it is
precisely the latter that consitute the curvature tensor $R$ at $x$.
Since the CEP involves the metric and its first derivatives, while its second
derivatives acquire a dynamical interpretation as the force-field of gravity,
one may regard CEP as a kinematical principle for the basic gravitational
potential variable $g_{\mu\nu}$ expressing a
`(gauge) possibility for local flatness' in GR by suitably
choosing the local frame (gauge) at $x$ to be inertial.}. It follows that the
symmetries of gravity are the isometries of $\cem$-localized; hence, one
arrives at a gauged or localized
version of the Lorentz group as the invariances of GR. This motivates us to formulate the Classical Local
Relativity Principle (CLRP) which, in a sense, is a corollary of the CEP above:

\noindent\underline{\it Classical Local Relativity Principle (CLRP)}: The group of
local (gauge) invariances of GR is isomorphic to the orthochronous
Lorentz group $L^{+}=SO(1,3)^{\uparrow}$ of symmetries of the Minkowski
space of SR.

In sum, the curved spacetime of GR may be modeled
after the locally Minkowskian tangent vector bundle
$TM:=\bigcup_{x\in X\subset M}\cem_{x}$, which is a sub-bundle of the dual of
the ${\bf G}$-bundle ${\cal{P}}$ mentioned in the introduction that consists of modules
of Cartan differentials\footnote{That is to say, Minkowskian covariant-tangent/cotangent
vectors which are dual to the Minkowskian contravariant
vectors in the fibers $\cem_{x}$ of $TM$. See also section 5.}and has as structure group
${\bf G}=SO(1,3)^{\uparrow}$, together
with a non-flat ${\bf g}=so(1,3)^{\uparrow}\simeq sl(2,\com)$-valued
spin-Lorentzian $\omg^{1}$-section $\cal{A}$.

Since, as it was mentioned in the introduction, causets effectively encode
the geometry of flat Minkowski space $\cem$, they can be thought of as local
kinematical structures representing the possible local causal relations in an
otherwise curved spacetime of events. The CEP, modified to fit a finitary,
curved and causal situation like ours, reads:

\noindent\underline{\it Finitary Equivalence Principle (FEP)}: a locally
finite curved causal space is
locally\footnote{Locality pending definition in our finitary context.}
a causet. Presumably, it is
the transitivity of causality as a partial order that must be renounced due
to gravity (Finkelstein, 1988, Raptis, 1998, 2000{\it
b,c})\footnote{Intuitively, gravity tilts the lightcone soldered at ({\it
ie}, with origin) each event, thus renders causality an intransitive relation
between them.}.

In other words, a curved smooth spacetime, as a causal space, is not globally transitive;
it is only locally (and kinematically) so\footnote{That is to say,
in the vertical
direction along each of the Minkowskian fibers of the curved co-vector bundle
${\cal{P}}$ above.}. Thus, the CEP
may be restated as a correspondence or reduction principle: as the
dynamically variable gravitational potential $g_{\mu\nu}$
reduces locally to the constant
$\eta_{\mu\nu}$ in GR, causality
becomes locally\footnote{As it was mentioned earlier, locality pending
definition in our finitary scenario (see the principle of Finitary Locality
below).} the constant
transitive partial order $\rightarrow$\footnote{With the CEP in mind, we may
call `$\rightarrow$' `the inertial Minkowskian causality'. In a curved causal
space causality only locally can be $\rightarrow$ (CEP).}. Equivalently, a
curved finitary causal space, one having a causal relation not fixed to a
globally transitive partial order, but with a dynamically variable local causality
between its events, is only locally reducible to a transitive, flat `inertial
causet'. Thus, as $\cem$ may be thought of as vertically extending, as an independent
kinematical fiber space, over every event of the curved smooth spacetime
manifold of GR, so an independent causet may be thought
of as being raised over every point-event of a curved finitary causal space.
Hence, the FEP almost mandates that a curved finitary causal space be
modeled after a finsheaf (or a bundle) of causets (over a finitary spacetime).
As a matter of fact, and also due to the finitary principle of Locality that
we will formulate shortly, we will see that a curved finitary causal space should
be modeled after a finsheaf of qausets (not of transitive causets) for
discrete locality's sake. Thus, some kind of `quantumness' will inevitably
be infused into our model of
the dynamics of finitary causality {\it ab initio}\footnote{This infusion converts our scheme
to `a model of the
dynamics of finitary {\it quantum} causality'. See our formulation of
the Finitary Local Superposition Principle below and the relevant discussion
in section 6.}. Before we give the
Finitary Locality Principle and its `corollary', the Finitary Local
Superposition Principle, we give the finitary analogue of the CLRP above:

\noindent\underline{\it Finitary Local Relativity Principle (FLRP)}: The
local invariance structure of a curved finitary causal space is a
finitary version of $L^{+}$. In a causal context, the work of Zeeman (1964,
1967) has shown that the symmetry structure $L^{+}$ of the flat Minkowski
continuum $\cem$, regarded as a causal space with
a causality relation between its events modeled after a (globally) inertial
partial order $\rightarrow$ which,
in turn, derives from $\cem$'s $\eta_{\mu\nu}$\footnote{See section 5 for
more on this.}, is isomorphic to the group
${\bf G}$ of causal automorphisms of $\cem$\footnote{Since the Alexandrov
causal topology of $\cem$ is defined by $\rightarrow$ (Alexandrov, 1956,
1967), it follows that ${\bf G}$ is the group of causal homeomorphisms of
$\cem$ (Torretti, 1981).}. In our case, and in view of the FEP, we infer
that a finitary version of $L^{+}$, which we call ${\bf G}_{n}$, comprises
the local relativity structure group of a curved finitary causal
space\footnote{Then, ${\bf G}_{n}$ consists of the local causal
homeomorphisms of the dynamically variable (and quantal) causal topology of a
curved finitary causal space. Since we plan to model the latter after
a finsheaf of qausets whose local structure is (by definition) the
qauset stalks over this curved finitary causal base space, ${\bf G}_{n}$ may
be equivalently thought of as consisting of the group of homomorphisms
(or automorphisms) of the quantal and causal
incidence algebras respresenting these qauset stalks (Raptis, 2000{\it b}). (See also remarks at the end of this section on the
significance of our choice to model the dynamics of a finitary quantum
causality by a finsheaf of qausets.)}. Now due to the local isomorphism
mentioned in the introduction between the Lie algebras $\ell^{+}=so(1,3)^{\uparrow}$ and
$sl(2,\com)$ in the
smooth ${\bf G}$-bundle ${\cal{P}}$, we may alternatively say that ${\bf
G}_{n}$ is the finitary version of the local relativity group $SL(2,\com)$
of GR in its spinorial gauge theoretic formulation
(Bergmann, 1957, Ashtekar, 1986, Baez and Muniain, 1994). A similar local relativity group for a curved finitary
quantum causal space was proposed in (Finkelstein, 1988)\footnote{We refer to
the local $SL_{2}$-invariances of the dyadic cell there.} and Selesnick (1994) found that $SL(2,\com)$ is the local relativity group for Finkelstein's reticular and curved quantum causal net.

In all the principles and remarks above, we mentioned the word `local'
without having transcribed the notion of classical locality to a curved finitary
causal scenario like ours. We do this now. The Classical Locality Principle
(CLP) in GR may be summed up to the following assumption:

\noindent\underline{\it Classical Locality Principle (CLP)}: The spacetime of
GR is modeled after a differential ($C^{\infty}$-smooth) manifold $M$
(Einstein, 1924)\footnote{Thus, the CLP may be viewed as
the requirement that all the dynamical laws of physics must be
differential equations, or more intuitively, that local dynamical actions
connect (influence) infinitesimally or `differentially' separated events
living in the tangent space at each event of the smooth spacetime continuum.
It follows that the CLP
requires physical observables or dynamical variables to be modeled after
(sections of) smooth differential forms (in ${\cal{P}}$), as mentioned in
the introduction. Thus, by `the local structure of the curved spacetime
manifold $M$' we mean `an event $x$ and the space of directions (contravariant
vectors) tangent to it' (Raptis and Zapatrin, 2000). In the bundle ${\cal{P}}$ this pertains to its,
Minkowskian by the CEP, fibers over each and every event $x$ of its base
space $M$.}.

Since a locally finite causal model like ours does not involve
(by definition) a continuous infinity of events like the $M$ above,
the CLP on $M$ may be translated in finitary causal terms to the following
requirement:

\noindent\underline{\it Definition of Finitary Locality (DFL)}:
In a causet, locality pertains to physical properties, to be interpreted as
observables or dynamical physical variables, with `effective range of action or
dynamical variation'
restricted to empty Alexandrov sets\footnote{See (Bombelli {\it et al.},
1987, Sorkin, 1990,
Raptis, 2000{\it b}) and the next footnote for a definition of these.}. Hence,
we shall demand that the following physical principle be obeyed by our
model of a curved finitary causal space:

\noindent\underline{\it Finitary Locality Principle (FLP)}: Dynamical
relations on a causet $(X,\rightarrow)$ involve only finitary local
observables\footnote{Thus, only dynamical changes of observables between
`nearest neighboring events' delimiting null Alexandrov sets in $X$ ({\it ie},
`$p,q\in X:\, (p\rightarrow q)\wedge(\not\!\exists
r:\, p\rightarrow r\rightarrow q)$', or in terms of the Alexandrov interval
bounded by $p$ and $q$, $A(p,q):=\{ r:~ p\rightarrow
r\rightarrow q\}=\emptyset$) are regarded as being physically
significant. This principle is an explication of the definition of
non-mediated (immediate) physical dynamical actions in the DFL above. Thus, by the DFL we anticipate the gravitational connection ${\cal{A}}$ in its finitary
and causal version ${\cal{A}}_{n}$, which is supposed to be the main
gravitational dynamical variable in our scheme, to be defined (as varying) on
such immediate causal arrows. See section 5 for more on this.}.

Some scholia on DFL and FLP are due here. Since in our reticular scheme we can
assume no dynamical properties varying between infinitesimally ({\it ie},
smoothly) separated events, we may as well define
local physical observables as the entities that vary between nearest
neighboring
events called `contiguous' from now on\footnote{In the last footnote, $p$ and
$q$ in the causet $(X,\rightarrow)$ are contiguous.}. The FLP
can be coined `the principle of contiguity in a finitary causal space' and it is
the reticular analogue of the CLP of GR, which,
in turn, as it was posited above, may be summarized to the assumption of a
$4$-dimensional differential
manifold model for spacetime (Einstein, 1924)\footnote{Parenthetically, we
mention that in this paper Einstein
concludes that the smooth geometrical manifold model for spacetime, which is
postulated up-front in GR for classical locality's sake, may
be thought of as an inert and absolute `aether'-like background structure on
which the whole theory of GR and the mathematical language
that supports it, classical Differential Geometry, is erected. In view of
his characteristic dissatisfaction with any theory that employs structures
that are absolute and non-dynamical, ultimately, `unobservable substances',
and in view of the reticular, molecular picture
of Nature that the quantum revolution brought about, we infer that Einstein
could not have been content with the smooth manifold model for spacetime
(Einstein, 1936, 1956). This is a significant aspect of our motivation to write the
present paper, for as it was mentioned in the introduction and it will
become transparent subsequently, one of the main
concerns of the present work is with finitary (coarse) and quantal
localizations of the
topological ({\it ie}, $C^{0}$) (Raptis, 2000{\it a}), as well as the
differential ({\it ie}, $C^{\infty}$-smooth) (Raptis and Zapatrin, 2000),
spacetime observables in a curved finitary causal and quantal situation. At least,
our central aim is to model `coarse (perturbing) quantal acts (operations) of
localization (local determination) of the main
gravitational observable ${\cal{A}}$'. See sections 5 and 6 for more
discussion of this.}. Also, by the FEP above, we expect that in a
curved finitary causal space gravity `cuts-off' the transitivity of
causality as a partial order and restricts the latter to empty Alexandrov
causal neighborhoods of contiguous events.

At this point it must be mentioned that the FLP, apart from
seeming rather natural to assume, was somehow `forced' on us by discrete
topological and local quantum causal considerations. In more
detail\footnote{For even more technical details and analytical discussion,
read the next section.}, it has been recently
shown (Breslav {\it et al.}, 1999) that the generating relation $\rho$ of the
Rota topology of the incidence algebra $\omg$ associated with a poset
finitary substitute $P$ of a continuous spacetime manifold as in (Sorkin, 1991)
is the same as the one generating the finite poset-topology of $P$ if and
only if one considers points in the Hasse
diagram of the latter that are immediately connected by the
partial order `$\rightarrow$' ({\it ie}, `contiguous events').
Then, if one interprets `$\rightarrow$' in
the finitary poset causally instead of topologically as in (Sorkin, 1995, Raptis,
2000{\it b}), and gives a cogent quantum interpretation to the structure of
the causal Rota algebra associated with it as in
(Raptis and Zapatrin, 2000, Raptis, 2000{\it b}), one is led to infer
that the physically significant, because local, causal connections between events
in a qauset are the contiguous, immediate ones; hence the FLP above. This was
first anticipated by Finkelstein (1988)\footnote{For
more on this demand for `local quantum causality' the reader is referred to
(Raptis, 2000{\it b,c}). We will return to it in the next three sections.}.
The FLP promotes this conjecture to a `physical axiom' (physical principle)
concerning the finitary dynamics of local (quantum) causality in a curved
locally finite causal space\footnote{To be represented by the finitary
connection ${\cal{A}}_{n}$ on the gauged finsheaf of qausets in section 5.}
in the same way that in the $C^{\infty}$-smooth spacetime $M$ of GR locality was
`forced' on Einstein by $M$'s own smoothness\footnote{In short, `the
physical law of classical gravity is mathematically modeled by a differential equation'.}.

According to the finitary principles formulated above, we may
say that in the same way that the CEP foreshadows a non-trivial connection
(and its associated curvature) in the smooth continuum-the main local dynamical
variable of GR as a gauge theory in ${\cal{P}}$, so FLP,
by `cutting-off' the transitivity property of
`$\rightarrow$' furnishes us with the crucial idea of how to model the
dynamics of a contiguous (local) quantum causality in a curved finitary
causal space, namely, one must define a non-trivial finitary connection (and its
associated curvature) on a finsheaf of qausets over it.
This connection, in turn, like its smooth spin-Lorentzian counterpart ${\cal{A}}$
on ${\cal{P}}$ over $M$ respects local relativistic causality\footnote{Since
it preserves the Minkowski lightcone soldered (with origin) at each
point-event
$x$ of $M$-the Minkowski lightcone in each fiber space $\cem_{x}$ of
${\cal{P}}$.}, should
somehow respect the local quantum causal connections in the
qauset fibers\footnote{That is to say, it should respect
the generating or `germ' relation
$\vec{\rho}$ of the Rota quantum causal topology of the qauset stalk of
the finsheaf in focus. We define $\vec{\rho}$ in the next section and germs
of (sections of) finsheaves of qausets in section 4.}. This highlights
and anticipates two very important
aspects of the present paper:

($a$) The finitary connection ${\cal{A}}_{n}$ (and its associated curvature)
derives from the local algebraic structure of the finsheaf of
qausets\footnote{That is to say, from the algebraic structure of the quantal and causal incidence
algebras-stalks of the finsheaf in focus.}. Thus, our scheme allows for a purely
algebraic and local definition of connection (and curvature) without reference
to a background geometric base space\footnote{This is in glaring contrast to
the situation in the curved geometrical manifold $M$
of GR where connection is identified with a parallel
transporter (of smooth tensor fields) along smooth finite spacetime curves, while
its associated curvature measures the anholonomy of such parallel transports
around smooth finite spacetime loops (G\"ockeler and Sch\"ucker, 1991).
Certainly,
both are non-local geometric conceptions of ${\cal{A}}$ and its ${\cal{F}}$.}
which will only serve as a surrogate host of ${\cal{A}}_{n}$ and which will
have to be
discarded, or at least be regarded as being physically insignificant,
at the quantal level, only to be recovered as a fixed inert
(non-dynamical) geometrical structure at the classical limit of an
inverse system of curved finsheaves of qausets\footnote{See discussion in
the introduction and sections 5, 6.}.

($b$) A sheaf (and a non-trivial connection on it) is the `right' ({\it ie},
the appropriate and natural) mathematical structure for modeling the dynamics ({\it ie}, the
curving) of local quantum causality, since, by definition, a sheaf is a local
homeomorphism (Bredon, 1967, Mallios, 1998, Raptis, 2000{\it a}), so that a
${\bf G}_{n}$-finsheaf of qausets by
definition respects the reticular local quantum causal topology of the qauset stalks,
while a non-flat ${\bf g}_{n}$-valued connection ${\cal{A}}_{n}$ on it
effectively encodes the `local twisting'
(curving) of these stalks, thus it represents the dynamics of a
locally finite quantum
causality. We will return to these issues more analytically in
the next three sections.

We close this section by giving the analogue of the kinematical Coherent Local Superposition Principle in (Finkelstein, 1988, 1991) for our finsheaves of qausets.

\noindent\underline{\it Finitary Local Superposition Principle (FLSP)}:
Stalk-wise in a finsheaf of qausets the latter superpose coherently. It
follows from the FLRP that the ${\bf g}_{n}$-valued connection
${\cal{A}}_{n}$ preserves this `stalk-wise quantum coherence of
qausets'\footnote{Since ${\cal{A}}_{n}$ takes values in the reticular (and quantal)
algebra ${\bf g}_{n}$ of Rota algebra homomorphisms which, in turn, by the
functorial equivalence between the category of finitary posets/poset morphisms
(or its corresponding category of
locally finite causets/causal morphisms) and the category of incidence Rota
algebras/Rota homomorphisms (or its corresponding category of qausets/qauset
homomorphisms) (Stanley, 1986, Raptis and Zapatrin, 2000, Zapatrin,
2000), it may be regarded as the
reticular and quantal version of Zeeman's (1964) Lie algebra $\ell^{+}$ of
orthochronous Lorentz transformations ({\it ie}, the infinitesimal causal
automorphisms) of the Minkowski continuum $\cem$ regarded as a poset causal space. We will return to this
remark in sections 4-6, but the upshot is that as a linear operator-valued map, ${\cal{A}}_{n}$ will preserve the local linear structure stalk-wise, hence, the local quantum coherence or quantum interference of qausets.}.

In the next section we present an algebraic approach to flat ({\it ie},
non-dynamical, non-gauged) local quantum causality, while in sections 4 and 5
we motivate the finsheaf theoretic point of view and we study a curved
principal finsheaf of qausets, respectively.

\section*{\normalsize\bf 3. FINITARY SUBSTITUTES AND THEIR FLAT QUANTUM
CAUSAL RELATIVES}

\indent\sloppy{In this section we motivate the modeling of qausets
after incidence algebras (Raptis, 2000{\it b}), so as to prepare the reader
for our representing the stalks of a finsheaf of qausets over some curved
finitary causal space as such Rota algebras in section 5. The relevance of
qauset theory to the problem of discrete Lorentzian (quantum)
gravity is also discussed. In particular, we approach the issue of
`discrete locality' or `finitary local causality' via qausets.
We quote the main result from
(Raptis, 2000{\it b}) that qausets are sound models of a local
and quantal version of the causets of (Bombelli {\it et al.}, 1987, Sorkin,
1990, 1995) and use it as a theoretical basis to implement the FLP
of the previous section, as well as to introduce the central physical
idea for curving local quantum causality in section 5 by localizing or
gauging a finsheaf of qausets (section 4), thus also realize the FEP of the
previous section.

The topological discretization of continuous spacetime (Sorkin, 1991)
has as its main aim the substitution of a continuum of events by some
finitary, but topologically equivalent, structure. The latter is seen to be
a $T_{0}$ poset. Such a finitary substitute for the continuous spacetime may
be viewed as an approximation of its continuous counterpart, but one
of physical significance, since it seems both theoretically and experimentally
lame to assume a continuum as a sound model of
what we actually experience ({\it ie}, record in the laboratory) as
`spacetime' (Raptis and Zapatrin, 2000).
The theoretical weakness of such
an assumption is the continuous infinity of events that one is in principle
able to pack into a finite spacetime volume resulting in the unphysical
infinities that plague classical and quantum field theory\footnote{See
opening quotation by Finkelstein (1991).}. The experimental
weakness of the continuous model of spacetime is that it
undermines the operational significance of our actual spacetime
experiments, namely, the fact that we record a finite number of events
during experimental operations of finite duration in laboratories of
finite size; altogether, in experiments of finite spatiotemporal
extent (Raptis and Zapatrin, 2000). Also, from a pragmatic point of
view, our localizations ({\it ie}, determinations of the loci) of
events are coarse or `approximate' and inflict uncontrollable
perturbations to the structure of spacetime\footnote{Even more so in
our scenario where spacetime is assumed to be fundamentally a quantum
system.}, thus our rough, because dynamically perturbing, measurements
of events may as well be represented by open sets about them (Sorkin,
1991, 1995, Breslav {\it et al.}, 1999, Raptis and Zapatrin, 2000,
Raptis, 2000{\it a}).

Of course, the discrete character of such finitary approximations of a
continuous spacetime ties well with the reticular and finite
characteristics that a cogent quantal description of spacetime
structure ought to have. Thus, if anything, topological
discretizations should prove useful in modeling the structure and
dynamics of spacetime at quantum scales (Raptis and Zapatrin,
2000). It must be stressed however that such a contribution to our
quest for a sound quantum theory of gravity is not mandatory from the
point of view of GR-the classical theory of gravity, since in the
latter the topology of spacetime is fixed to that of a locally
Euclidean manifold\footnote{That is to say, the spacetime of GR is
assumed to be locally homeomorphic to the `frozen' ({\it ie}, non-varying)
Euclidean continuum $\real^{4}$ (see opening quotation).}, while only
the Lorentzian metric on it is assumed to be a dynamically variable
entity. Effectively, $g_{\mu\nu}$ is the sole `observable' in
GR. However, it seems rather {\it ad hoc} and unreasonably limited in
view of the persisting and pestilential problem of the quantum
localization of spacetime events to assume that only the metric, but
not the topological structure of the world, is subject to (quantum)
dynamical fluctuations and variations. Such a theory of `spacetime
foam', that is to say, of a dynamically fluctuating quantum spacetime
topology, has been aired for quite some time now (Wheeler, 1964), and
it is akin in spirit to the topological discretizations developed in
(Sorkin, 1991), as well as to their quantal relatives in (Raptis and
Zapatrin, 2000)\footnote{See (Sorkin, 1995) for some discussion on
this affinity.}.

On the other hand, in view of the unphysical, non-dynamical,
non-relativistic, space-like nature of the constant, two-way, spatial
connections between events that define the (locally) Euclidean
topology of the classical spacetime continuum $M$, there is an
important affinity between our quest for a dynamical theory of (local)
quantum causal topology and the problem of constructing a reasonable
quantum theory of gravity. To understand this close relationship, we
must change focus of enquiry from a theory of spatial Euclidean
connections between points\footnote{Ultimately, a `topo-' or
`choro-logy' (Greek for `a theory of space').} to a more physical,
because relativistic, temporal or causal spacetime
topology\footnote{Ultimately, a `chrono-logy' (Greek for `a theory of
time').} between events as the quotation opening this paper and the
following one from (Finkelstein, 1988) suggest:

\begin{quotation}

{\footnotesize It is therefore crucial to make use of the proper
physical topology. The usual combinatory topology is founded on a
symmetric concept of connection derived from experience with
Riemannian and ultimately Euclidean geometry. It assumes the existence
of spatial connections and puts them on the same footing as timelike
ones, when (in the absence of any signs of tachyons) it is quite
doubtful that either exist at all. A relativistic topology should deal
with causal connections among events, not spatial connections among
objects. At the continuum level, the Alexandroff\footnote{This is the
A. D. Alexandrov (1956, 1967) (our footnote).} topology is already
suitably relativistic. It is thus only necessary to construct a
relativistic discrete topology and homology theory on the basis of the
causal connection ${\bf c}$\footnote{`${\bf c}$' in (Finkelstein,
1988) stands for an atomistic, local, dynamical relation that defines
a dynamically variable causal topology between events. Subsequent
`algebraization' of ${\bf c}$ results in a model for the dynamics of
local quantum causality, called `the quantum causal net', which is
quite similar to the one that we propose here. See also (Raptis, 2000{\it
b}).}.}

\end{quotation}

As it was mentioned in the previous section, in GR, the gravitational
potential, which is identified with the metric $g_{\mu\nu}$ of
spacetime, may also be thought of as encoding complete information
about the local causal relations between events\footnote{That is to
say, at every event $x$, $g_{\mu\nu}(x)$, which can be reduced to the
Minkowski metric $\eta_{\mu\nu}$ by the CEP, delimits the lightcone at
$x$, which in turn determines the possible (kinematical) causal
relations between it and its local neighbors in the Minkowski space
tangent to it (`infinitesimal locality' or `local causality'). Albeit,
as Finkelstein successfully observed in (1988), Einstein, following
Riemann, `metrized' causality in GR instead of `topologizing'
it.}. Thus, GR may also be interpreted as the dynamical theory of `the field
of local causality'. It follows that a quantum theoresis of the dynamics of
causal connections between spacetime events may lead to, if not
just give us invaluable clues about, a `classically conceived
quantum theory of gravity'-`the quantization of the gravitational field
$g_{\mu\nu}(x)$ of GR'\footnote{Otherwise known as
`Quantum General Relativity'. Again,
it is Finkelstein who successfully observed that gravity is
of an essentially local causal-topological nature (Finkelstein, 1988).}. In
short, there
probably is a way from a dynamical theory of local quantum causality to the
graviton, but not the other way around\footnote{Read the end of the quotation
from Bombelli {\it et al.} (1987) given in the introduction and (Sorkin,
1990, 1995).}. A full-fledged non-commutative topology for curved ({\it ie}, dynamical) local quantum causality will be rigorously formulated in the scheme theoretic language of modern algebraic geometry and its categorical outgrowth, topos theory, in a coming paper (Raptis, 2000{\it e}).

However, it must be stressed that it is quite clear, at least from a `gedanken-experimental' point of view,
why GR and Quantum Theory are incompatible: the more
accurately one may try to determine the spacetime metric\footnote{That is
to say, `localize' it.}, the more energy one must employ,
the stronger the dynamical perturbations inflicted
on it, the higher the uncertainty of its local
determination\footnote{That is to say,
the CEP on which GR is based comes straight into conflict
with the Uncertainty Principle on which Quantum Theory is founded (Candelas
and Sciama, 1983, Donoghue {\it et al.}, 1984, 1985).}. Another way to
say this is that we can not distinguish
spacetime events\footnote{Or equivalently, measure the
distance of their separation via the gravitational potential $g_{\mu\nu}$.}
at a resolution higher than the Planck length ($l_{P}\approx 10^{-33}cm$)
without creating a black hole
which, in return, `fuzzies' their separation in some sense\footnote{This is
the quintessential paradox of event-localization that makes the conception of a
quantum theory of gravity hard even in principle: the
more accurately we try to localize spacetime events, the more we blur them,
so that our sharpest determinations of them can be modeled after
coarse, rough, fuzzy, `dynamically fluctuating' open neighborhoods about
them as in (Sorkin, 1991, 1995, Zapatrin, 1998, Breslav {\it et al.}, 1999, Raptis and Zapatrin, 2000, Raptis, 2000{\it a}).}. This limitation
alone is sufficient to motivate some kind of `topological foam' conception of spacetime at
quantum scales (Wheeler, 1964).
An analogous incompatibility (of physical principles) that may hinder the
development of a quantum theory of the dynamics of a finitary causality
has not been predicted yet. We hope that such a fundamental
conflict of physical principles will be absent {\it ab initio} from an
innately locally finite dynamical theory of quantum causality\footnote{In
other words, a `dynamical finitary quantum causal topology'.}, or at least from the relevant kinematics for such a theory, like the one
that we will propose in section 5.

This lengthy prolegomenon to the introduction of the essentially flat qausets
in (Raptis, 2000{\it b}) highlights two important aspects
of our present endeavor: ($a$) our locally finite, and
subsequently to be gauged, qausets, may evade {\it ab initio}
the infinities of Quantum GR on a smooth manifold, and
($b$) as quantum causal-topological structures, they grapple with
the problem of the structure and dynamics of spacetime at quantum scales at a
level deeper than Quantum GR-proper which is supposed to
study the quantum aspects of the dynamics of the metrical structure of the
world, because we have seen already at the classical level that causality, as
a partial order, and its morphisms, determine the geometric structure
of flat Minkowski space and its symmetries (Robb, 1914, Alexandrov, 1956,
1967, Zeeman, 1964, 1967, Bombelli and Meyer, 1989, Sorkin, 1990).
Afterall, as Bombelli {\it et al.} (1987) successfully observed in the
excerpt presented in the introduction, it is
such a model for events and their causal relations that uniquely
determines spacetime as a $4$-dimensional, continuous ($C^{0}$),
differential ($C^{\infty}$-smooth) and Lorentzian metric ({\it ie}, of signature $\pm2$) manifold.

We commence our brief review of qausets by first presenting elements
of finitary substitutes for continuous spacetime topology (Sorkin, 1991,
Raptis and Zapatrin, 2000, Raptis, 2000{\it a,b}). Let $X$ be a bounded region in a
continuous spacetime manifold $M$\footnote{By `bounded' we mean `relatively
compact' ({\it ie}, a region whose closure is compact). By `continuous' we
mean the $C^{0}$ aspects of classical spacetime ({\it ie}, spacetime as a topological
manifold).} and ${\cal{U}}=\{ U\}$ a locally finite open cover of
it\footnote{That is to say, every point-event $x$ in $X$ has an open neighborhood
$O(x)$ that meets only a finite number of open sets $U$ in
$\cal{U}$.}. The boundedness of $X$ stands for the pragmatic restriction of
our experimental discourse with spacetime in laboratories of finite
spatio-temporal extent; while, its locally finite open covering by $\cal{U}$
reflects the fact that within this experimental activity of finite duration
and spatial extension `for all practical purposes' we record a finite number
of events coarsely or `rougly', that is to say, by determining a finite number of open
sets about them (Breslav {\it et al.}, 1999, Raptis and Zapatrin, 2000, Raptis,
2000{\it a}).

Any two points $x$ and $y$ of $X$ are indistinguishable with respect to its
locally finite open cover ${\cal{U}}$ if $\forall U\in
{\cal{U}}:\, x\in U\Leftrightarrow y\in U$. Indistinguishability with
respect to the subtopology ${\cal{T}}({\cal{U}})$\footnote{${\cal{T}}$
consists of arbitrary unions of finite intersections of the open sets in
${\cal{U}}$.} of $X$ is an equivalence relation on the latter's points and
is symbolized by $\stackrel{{\cal{U}}}{\sim}$. Taking
the quotient $X/\stackrel{{\cal{U}}}{\sim}=:F$ results in the substitution
of $X$ by a space $F$ consisting of equivalence classes
of its points, whereby two points in the same equivalence class are covered by
({\it ie}, belong to) the same,
finite in number, open neighborhoods $U$ of ${\cal{U}}$, thus are
indistinguishable by (our coarse observations in) it.

Let $x$ and $y$ be points belonging to two distinct equivalence classes
in $F$. Consider the smallest
open sets in the subtopology ${\cal{T}}({\cal{U}})$ of $X$
containing $x$ and $y$ respectively given by:
$\Lambda(x):=\cap\{ U\in {\cal{U}}:\,
x\in U\}$ and $\Lambda(y):=\cap\{ U\in {\cal{U}}:\, y\in U\}$.
Define the relation $\rightarrow$ between
$x$ and $y$ as follows: $x\rightarrow y\Leftrightarrow
\Lambda(x)\subset\Lambda(y)\Leftrightarrow x\in\Lambda(y)$.
Then assume that $x\stackrel{{\cal{U}}}{\sim}y$ in the previous
paragraph stands for $x\rightarrow y$ and $y\rightarrow
x$\footnote{That is to say, $x$ and $y$ have the same smallest open
neighborhood about them with respect to ${\cal{T}}({\cal{U}})$.}.
$\rightarrow$ is a partial order on $F$ and the continuous $X$ has been
effectively substituted by the
finitary $F$ which is a $T_{0}$ topological space having the structure
of a poset (Sorkin, 1991). Sorkin uses the
finitary topological and partial order theoretic languages interchangeably exactly due to this
equivalence between $T_{0}$ finitary substitutes and posets. For future
purposes we distill
this to the following statement: in (Sorkin, 1991) a partial order is
interpreted topologically. We call it `topological partial
order' and the poset encoding it `topological poset' (Raptis, 2000{\it b}).

Topological posets have an equivalent representation as simplicial complexes
(Raptis and Zapatrin, 2000). One may represent a finitary spacetime
substitute by a simplicial complex by considering the so-called nerves
of locally finite open covers of $X$ (Alexandrov, 1956, 1961)\footnote{This
is the P. S. Alexandrov.}.
We may recall that the nerve ${\cal{N}}$ of a covering
$\cal{U}$ of a manifold $X$ is the simplicial complex whose
vertices are the elements of $\cal{U}$ and whose simplices are formed
according to the following rule. A set of vertices, that is to say,
elements of the locally finite covering $\{U_0, \ldots , U_k\}$ form a
$k$-simplex of ${\cal{N}}$ if and only if they have nonempty
intersection:

\[
\{U_0, \ldots , U_k\} \in {\cal{N}} \: \Leftrightarrow \:
U_0 \cap U_1 \cap \ldots \cap U_k \neq \emptyset
\]

\noindent Any nerve ${\cal{N}}$, being a simplex, can be equivalently treated
as a poset, denoted also by ${\cal{N}}$. The points of the poset ${\cal{N}}$
are the simplices of the complex ${\cal{N}}$, and the arrows are drawn
according to the rule: \[
p \to q \: \Leftrightarrow \: p \mbox{ is a face of } q.
\] \medskip

\noindent In the nondegenerate cases, the posets associated with Alexandrov nerves and
those produced by Sorkin's `equivalence algorithm' yielding $F$ from $X$
relative to $\cal{U}$ as described above are the
same, so that both are `topological posets' according to our denomination
of $F$.

In (Raptis and Zapatrin, 2000) an algebraic representation of topological
posets was presented using the so-called Rota incidence algebras associated with
posets (Rota, 1968). The Rota incidence algebra $\omg$ of a poset $P$ was
defined there by using Dirac's quantal ket-bra notation as follows

\[
\omg(P)=\spn\{\ketbra{p}{q}:\, p\rightarrow q\in P\} ,
\]

\noindent with product between two of its ket-bras given by

\[
\ketbra{p}{q} \cdot \ketbra{r}{s} =
\ket{p} \braket{q}{r} \bra{s} =
\braket{q}{r} \cdot \ketbra{p}{s} =
\left\lbrace \begin{array}{rcl}
\ketbra{p}{s} &,& \mbox{if } q=r \cr
0 && \mbox{otherwise.}
\end{array} \right.
\]

\noindent Evidently, for the definition of the product in $\omg$, the
transitivity of the
partial order $\rightarrow$ in $P$ is used. $\omg(P)$, defined thus, is
straightforwardly verified to be an associative
algebra\footnote{The associativity of the product of
the incidence algebra
$\omg$ is due to the transitivity of the partial order
$\rightarrow$ of its associated
poset $P$. As we saw in section 2, it is precisely the
latter property of causality, when modelled after
$\rightarrow$, that is responsible for the (global)
flatness of Minkowski space determined by $\rightarrow$.
It follows that a localization or gauging of causets
and their corresponding qausets in order to curve them,
by providing a connection on a principal finsheaf of
theirs, will `cut-off' the transitivity of the causets
and the associativity of their corresponding qausets,
and will restrict it locally ({\it ie}, stalk-wise) in
the finsheaf (section 5) thus implement the FEP of
section 2. We will briefly return to this conception of
global non-associativity in a curved topos of finsheaves
of qausets in $(f^{'})$ of section 6.}. When $P$ is a finitary topological poset
in the sense of Sorkin
(1991), its associated incidence algebra is called
`topological incidence
algebra' (Raptis, 2000{\it b}).

We may define purely algebraically a topology on any incidence algebra
$\omg$ associated with a poset $P$ by
considering its primitive spectrum $\cal{S}$ consisting of the kernels of
its irreducible representations, which are primitive ideals in it, in the
following way according to Breslav {\it et al.} (1999): with every
point $p$ in $P$ the following ideal in $\omg$ is defined

\[
I_{p}=\spn\{\ketbra{q}{r} :\, \ketbra{q}{r}\not= \ketbra{p}{p}\} ,
\]

\noindent so that the Rota topology of $\omg(P)$ is generated by the
following relation $\rho$ between `points' $I_{p}$ and $I_{q}$ in its
primitive spectrum $\cal{S}$

\[
I_{p}\rho I_{q}\Leftrightarrow I_{p}I_{q}(\not= I_{q}I_{p})\stackrel{\not=}
{\subset} I_{p}\cap I_{q}.
\]  \noindent It has been shown that the Sorkin topology of a topological poset
$P$ is the same as the Rota topology of its associated topological incidence
algebra $\omg(P)$
exactly when the generating relation $\rho$ for the latter is the transitive
reduction $\imm$ of the partial order arrows $\rightarrow$ in $P$
(Breslav {\it et al.}, 1999, Raptis, 2000{\it b})\footnote{That is to say, $I_{p}$ is $\rho$-related to $I_{q}$
if and only if $(p\imm q)\Leftrightarrow[(p\rightarrow q)\wedge(\not\exists r:~ p\rightarrow
r\rightarrow q);~ p,q,r\in P]$ ({\it ie}, only for
immediately connected contiguous
vertices in $P$).}. This means essentially that the
`germ-relations' for the Rota topology on the algebra
$\omg$ associated with the finitary topology $P$ are precisely the
immediate arrows $\imm$ in the latter topological poset. This is an important
observation to be used shortly in order to define in a similar way the germs
of quantum causal relations in a qauset with respect to which
finsheaves of the latter will be defined in the next section as structures
that preserve these local quantum causal topological
`germ relations'\footnote{That is to say, as `local
quantum causal homeomorphisms'. In (Raptis, 2000{\it e}) this `topologization' of the primitive spectra of incidence Rota algebras associated with Bombelli {\it et al.}'s causets, that is, the qausets in (Raptis, 2000{\it b}), will be the chief motivation for the definition of primitive finitary spacetime schemes (finschemes) of qausets and to study their localization properties, their non-commutative topological aspects, as well as the organization of these finschemes into a topos-like structure called `quantum topos' (see also $f^{'}$ in section 6).}.

To this end, we give the definition of qausets. A causet
is defined in (Bombelli {\it et al.}, 1987) as ``{\it a locally finite set of points
endowed with a partial order corresponding to the macroscopic relation that defines
past and future}". Local finiteness may be defined as
follows: use $\rightarrow$ of a poset $P$, interpreted now as a causal relation on the causet,
to redefine $\Lambda(x)$ for some $x\in P$ as
$\Lambda(x)=\{ y\in P:\, y\rightarrow x\}$,
and dually $V(x)=\{ y\in P:\, x\rightarrow y\}$. $\Lambda(x)$ is the
`causal past' of the event $x$, while
$V(x)$ its `causal future'. Then, local finiteness requires the so-called
Alexandrov set $V(x)\cap \Lambda(y)$ to be finite for all
$x,y\in P$ such that
$x\in \Lambda(y)$. In other words, only a finite number of events
`causally mediate' between any two events
$x$ and $y$, with $x\rightarrow y$, of the causet $P$. In a sense, the
finitarity of the topological posets translates
by Sorkin's semantic switch to the local finiteness of causal sets, although
it must be stressed that the physical theories that they support,
the discretization of topological manifolds in (Sorkin, 1991) and causet theory in
(Bombelli {\it et al.}, 1987 and Sorkin, 1995) respectively,
are quite different in motivation, scope and aim (Sorkin, 1990, 1991, 1995, Raptis,
2000{\it b})\footnote{See also discussion in $(d^{'})$ of section 6.}.

On the other hand, it was Sorkin who first insisted on a change of physical
interpretation for the partial order $\rightarrow$ of finitary posets $P$
``{\it from a relation encoding topological information about bounded regions of continuous
spacetimes, to one that stands for the relation of causal succession between
spacetime events}" (Sorkin, 1995). The following quotation from that paper is very
telling indeed\footnote{This quotation can be also found in (Raptis, 2000{\it b}).}:

\begin{quotation}
{\footnotesize Still, the order inhering in the finite topological space seemed to be very
different from the so-called causal order defining past and future. It had
only a topological meaning but not (directly anyway) a causal one.
In fact the big problem with the finite topological space was that it seemed to
lack the information which would allow it to give rise to the continuum in all its aspects, not just in the
topological aspect, but with its metrical (and therefore its causal) properties as well...The way out of the
impasse involved a conceptual jump in which the formal mathematical structure remained constant, but its physical
interpretation changed from a topological to a causal one...The essential realization then was that, although order
interpreted as topology seemed to lack the metric information needed to
describe gravity, the very same order
reinterpreted as a causal relationship, did possess information in a quite straightforward sense...In fact it took me
several years to give up the idea of order-as-topology and adopt the causal set alternative as the one I had been
searching for...}
\end{quotation}

In (Raptis, 2000{\it b}), this `semantic switch' was
evoked to reinterpret the incidence algebras associated with the finitary posets
in (Raptis and Zapatrin, 2000)
from topological to causal. Thus, causal incidence algebras were defined as
the $\omg$s associated with finitary posets $P$ when the latter are
interpreted as causets {\it \`{a} la} Bombelli {\it et al.} (1987).
Of course, in our
pursuit of a cogent quantum theory of the dynamics of causality and, {\it in
extenso}, of gravity, such a change of physical meaning of finitary partial
orders from spatial/choro-logical/topo-logical to
temporal/chrono-logical/causal is very
welcome for the reasons given earlier in this section.

Finally, in (Raptis, 2000{\it b}) the quantum physical interpretation given
to topological incidence algebras in (Raptis and Zapatrin, 2000) was
also given directly to causal incidence algebras. Thus, qausets were
defined as the causally and quantally interpreted incidence algebras
associated with poset finitary substitutes of continuous
spacetimes\footnote{Thus, unlike Regge
(1961) who metrized the simplicial skeletons of spacetime and
arrived at a simplicial gravity in the same spirit as to how Einstein
identified the gravitational potential with the spacetime metric in the
continuum theory, we, following Sorkin (1995), causalize the simplicial-poset
topological discretizations of spacetime in (Alexandrov, 1956, 1961, Sorkin,
1991) and their quantum relatives in (Raptis and Zapatrin, 2000) with an aim to
arrive at a finitary quantum causal version of gravity. Einstein and Regge
metrized topology,
we causalize it ({\it ie}, we re-interpret topology causally), and subsequently
(section 5) we regard it as a dynamical local variable as the opening quotation from
Finkelstein (1991) motivated us to).}. It
follows that the generator $\imm$ of
topological relations in the topological
posets of Sorkin becomes the germ $\vec{\rho}$ of quantum causal relations
in qausets\footnote{Due to its causal instead of topological meaning, we
are going to write $\vec{\rho}$ instead of $\rho$ from now on for the local quantum causal topological variable.}. Its
interpretation is as `immediate quantum
causality'\footnote{The epithet `quantum' refers precisely to the possibility
for coherent quantum superpositions of the causal arrows of $P$ in its
associated incidence algebra $\omg(P)$ (Raptis and Zapatrin, 2000, Raptis,
2000{\it b}).} and it is exactly due to its natural
Rota-algebraic representation that qausets are (operationally) sound models
of quantum causal spaces (Finkelstein, 1988, Raptis, 2000{\it b}).
$\vec{\rho}$ is the algebraic
correspondent in the causal incidence algebra $\amg$ of the immediate causal
relation $\imm$ of its associated causet $\vec{P}$\footnote{Notice again the
arrow over $P$ when the latter is interpreted as a causet rather than as a
topological poset.}.

The immediate quantum causality represented by $\vec{\rho}$ in the
incidence algebra associated with a causet is ideal for implementing the
FLP of the previous section. In particular, it
explicitly shows that the physically significant, because local, quantum
causality is the relation $\imm$ between immediately separated events in a
finitary
spacetime $X$ (Finkelstein, 1988, Raptis, 2000{\it b}). Its non-local
transitive closure, the partial order $\rightarrow$ in the associated
causet $\vec{P}$, generates $\vec{P}$'s (globally) inertial Minkowskian
causal topology which, being a finitary poset, essentially
determines a locally finite version of flat Minkowski space and its global
orthochronous Lorentz symmetries (Robb, 1914, Alexandrov, 1956, 1967,
Zeeman, 1964, 1967).

It follows that in order to curve qausets, a gauged or localized
version of $\vec{\rho}$ must be employed, that is to say, we should
consider a dynamical local quantum causal connection relation that only locally
({\it ie}, event-wise)\footnote{As we said in section 2 when we discussed
the FLP, in the finsheaf theoretic context of
section 5 locality will correspond to `stalk-wise over an event in a curved
finitary causal space'.} reduces
to a transitive partial order-the inertial Minkowskian causality of a
reticular and quantal Minkowski space (as a qauset) according to
the FEP. In turn, this means that only the transitive reduction $\imm$ of the
flat inertial causality $\rightarrow$ will be `operative' ({\it ie},
physically significant) in a curved finitary quantum causal space. We will model
this conjecture by a non-flat connection ${\cal{A}}_{n}$ on a finsheaf of qausets
in section 5. In a sense, such a non-trivial
connection ${\cal{A}}_{n}$ will be seen to `cut-off' the transitivity of causality
as a partial order $\rightarrow$, so as to
restrict the latter to immediate causal neighborhoods of events defined
by $\imm$, thus respect the FLP\footnote{Hence, ${\cal{A}}_{n}$ will also qualify
as a finitary (and quantal) local dynamical variable ({\it ie}, `observable')
in the aforementioned gauged finsheaf of qausets. ${\cal{A}}_{n}$'s `purely
algebraic character' mentioned before ({\it ie}, its essentially
non-commitment to
any geometric background base spacetime whatsoever) is that it represents
the dynamics of the local quantum causal topology
$\vec{\rho}$ of the quaset stalks of the finsheaf and the latter is purely
algebraically defined in the spectra of these stalks (see section 5). The purely algebraico-categorical nature of ${\cal{A}}_{n}$ will be also pronounced in our scheme theoretic presentation of qausets in (Raptis, 2000{\it e}).}.

We conclude the present section by discussing briefly two relatively
important aspects of qausets, one physical, the other
mathematical.

The physical aspect of qausets pertains to their operational
significance\footnote{Here, we refer to the classical definition of
operationality by Bridgman (1936) according to which: ``{\it in dealing with
physical situations, the operations that give meaning to our physical
concepts should properly be physical operations, actually carried out}".}.
While the operational soundness of quantum discretized
spacetimes has been fairly established (Raptis and Zapatrin, 2000) in that
we have a sound operational interpretation of quantal topological
incidence algebras\footnote{Briefly, the operational meaning of quantal
incidence algebras goes as follows: the finitary substitutes of continuous
spacetime, be it posets as in Sorkin (1991) or simplicial complexes as in
Alexandrov (1956, 1961), stand for coarse ({\it ie}, approximate) determinations
of the loci of spacetime events. Since these measurements are imperfect and,
in a quantum sense, they inflict uncontrollable perturbations on spacetime,
what they actually determine is `rough', `blurry' or `fuzzy' regions about
events that can
be modeled after open sets. The latter are supposed to model the irreducible
quantum uncertainties of space-time localizations that are expected at scales
smaller than
Planck's $l_{P}\approx 10^{-33}cm$ and $t_{P}\approx 10^{-42}s$ (Capozziello
{\it et al.}, 2000). By covering the region of spacetime under
experimental focus by a finite number of such neighborhoods, and by keeping
track of the mutual intersections of these coarse
determinations ({\it ie}, their nerves), enables one to build the Alexandrov
simplicial complexes of such an experimental activity, or their topologically equivalent
finitary posets of Sorkin (Raptis and Zapatrin, 2000). This is a formal
account of ``{\it what we actually
do to produce spacetime by our measurements}" (Sorkin, 1995)-the essence
of operationality. In the context of quantum topological incidence algebras,
sound operational meaning has been given to the corresponding
algebraic structures and it all is operationally sound (Raptis and Zapatrin,
2000). In the latter reference, Heisenberg's algebraic
approach to quantum theory, that
points to a more general and abstract conception of operationality where
quantum physical operations are organized into an algebra (call it
`quantum operationality' to distinguish it from Bridgman's classical
conception of this notion that is not based on a specific mathematical
modeling of the actual physical operations), and according to which
``the physical operations that give meaning to our conceptions about the
quantum realm, ultimately, our
own experimental actions on the quantum system that define
({\it ie}, prepare and register (Finkelstein, 1996)) its observable
quantitites, should be modeled algebraically", was used to interpret quantum physically the incidence algebras
associated with Sorkin's (1991) finitary topological posets.}, we still
lack such an account for qausets.

Now, GR's operational significance can be summarized in the
following: $g_{\mu\nu}(x)$, which mathematically represents the local
gravitational potential, is supposed to encode all the information about
our local experimental tampering with spacetime events via synchronized clocks and
equicalibrated rulers, so that, in principle, from the data of such a local experimental
activity, one can construct the metric tensor at a neighborhood of an event
(the latter serving as the origin of our laboratory/frame). In such an
operational account, there is little room left for a `passive' realistic interpretation
of the gravitational field as an independent entity or substance `out there' whose
interaction with our instruments yields readings of events. The operational
approach is in an important sense more active in that it entails that
spacetime attributes are extracted from `it'\footnote{`It' referring to the
physical system `spacetime'.} by our very experimental actions on
({\it ie}, our planned, coordinated and controlled observations of) `it'.
Also, this seems
to be more in accord with the observer-dependent conception of physical reality
that Quantum Theory supports (Finkelstein, 1996)\footnote{The reader is
referred to that book for a deep operational, in fact, pragmatic,
`unification' of the basic principles of relativity and quantum theory.
Finkelstein elevates the aforementioned `quantum operationalism' of
Heisenberg, to a philosophically sound `quantum relativistic pragmatism' which, in
a nutshell, holds that our rationalization about the quantum and
relativistic realm of Nature and
the properties of relativistic quantum systems-our
quantum relativistic calculus (logic or set theory) so to speak, is defined by our
very experimental (inter)actions on (with), ultimately, by our dynamical
transformations (in the general operational sense of actions of preparation, propagation and registration) of, relativistic quanta.}.

For the causets of Bombelli {\it et al.} (1987) and Sorkin (1990), Sorkin (1995) contended
that an operational interpretation is rather unnatural and lame. On the other hand,
in view of the algebraic structure of qausets and the sound
quantum-operational interpretation {\it \`a la} Raptis and Zapatrin (2000)
that their topological counterparts were given,
and because as we mentioned in section 2 the local field of gravity
$g_{\mu\nu}(x)$ can also be interpreted as the dynamical field of the local
causal topology of spacetime, we still hope for a sound operational
interpretation of them. At the
end of the next section we present our first attempt at a sound operational
interpretation of the locally finite quantum causality encoded in qausets
based on the analogous operational
meaning of the finitary poset substitutes of continuous spacetimes and their
incidence algebras (Sorkin, 1991, 1995, Raptis and Zapatrin, 2000). A more
thorough presentation of the operational character of qausets will be given
in a coming paper (Raptis, 2000{\it d}).

The mathematical aspect of qausets that we would like to discuss next
is their differential structure. Recently, there has been vigorous research
activity on studying differential calculi on finite sets and the
dynamics of networks (Dimakis {\it et al.}, 1995), as well as on defining
some kind of discrete
Riemannian geometry on them (Dimakis and M\"{u}ller-Hoissen, 1999).
The main result
of such investigations is that with every directed graph a discrete
differential calculus may be associated. It follows that for the locally
finite posets underlying qausets $\amg$ ({\it ie}, the causets $\vec{P}$
associated with them), which are also
(finitary) digraphs, there is a discrete differential calculus associated
with them (Raptis and Zapatrin, 2000, Zapatrin, 2000). In this sense, but from a
discrete perspective, a partial order determines not only
the topological ($C^{0}$), but also the differential ($C^{\infty}$)
structure of the spacetime manifold with
respect to which the Lorentzian $g_{\mu\nu}$, which is also determined
by causality as a partial order\footnote{At least locally in a curved
spacetime (see section 2).}, is then defined as a smooth
field\footnote{See the quotation from (Bombelli {\it et al.}, 1987) in the
introduction.}.

However, as we noted in the introduction, the K\"ahler-Cartan type of
discrete differential operator $\ddd$ defined in such calculi on
finite sets is a flat sort of connection (Mallios, 1998). This is not surprising,
since the underlying finite space(time) $X$ is taken to be a structureless
point-set\footnote{In a sense, a kind of disconnected, non-interacting
`dust'.}. All the digraphs supporting
such calculi are assumed to be transitive, so that if some causal
interpretation was given to their arrows, by our heuristic principles of
section 2 concerning the
relation between an inertial transitive causality and flatness, their
corresponding differential calculi should be flat as well\footnote{That is to
say, the differential operators defining such calculi are flat connections
in the sense of Mallios (1998).}. This is the `curvature problem' alluded to
in the introduction. To evade it, in section 5 we straightforwardly gauge (a finsheaf of) qausets
so that a non-flat connection ${\cal{A}}_{n}$ is naturally defined on them. The
physical interpretation of such a gauging
of the $\ddd$ of flat qausets to the $D=\ddd+{\cal{A}}$
of the curved finsheaf of qausets, will be the
first essential step towards a finitary, causal and quantal version of
Lorentzian gravity.

In the next section we recall the
finsheaves from (Raptis, 2000{\it a}). Our principal
aim is to review the sense in which a finsheaf of continuous maps over
Sorkin's topological posets approximates the sheaf of $C^{0}$-topological observables
over a continuous spacetime manifold, then try to `read' a similar physical
meaning for a finsheaf of qausets, namely,
that they model finitary and quantal replacements
of the causal relations between events in a bounded region of flat
Minkowski space $\cem$, as well as the causal nexus of $C^{\infty}$-smooth
fields in such a region of the smooth differential manifold $\cem$.
At the same time,
the finsheaves of their continuous symmetries may be thought
of as locally finite substitutes of the continuous orthochronous Lorentz topological
($C^{0}$) Lie group manifold $SO(1,3)^{\uparrow}$. In such
a scenario, not only the operational significance of our own coarse
`approximations'
of spacetime structure and its dynamics will be highlighted, but also the
operational meaning of our rough and dynamically perturbing determinations of its symmetries.

\section*{\normalsize\bf 4. FINITARY SPACETIME SHEAVES AND THEIR FLAT QUANTUM
CAUSAL DESCENDANTS}

\indent\sloppy In (Raptis, 2000{\it a}), a finsheaf $S_{n}$
of continuous
functions on a bounded region $X$ of a topological spacetime manifold $M$
was defined as the sheaf of sections of continuous maps on $X$ relative
to its covering by a locally finite collection of open subsets of $M$. Since,
as we saw in the previous section, for every such finitary open cover
${\cal{U}}_{n}$ of $X$ a finitary topological poset $F_{n}$
was defined and seen to effectively substitute $X$,
the aforementioned sheaf can be thought of as having $F_{n}$ as
base space. Thus, we write $S_{n}(F_{n})$ for such a finsheaf
(Raptis, 2000{\it a}). Indeed, $S_{n}$ was seen to have locally the same
finite poset-topology as its base space $F_{n}$\footnote{Technically speaking,
$S_{n}$ is locally
homeomorphic to the finite topological space $F_{n}$ of (Sorkin, 1991).} (Raptis, 2000{\it a}),
hence its qualification as a sheaf (Bredon, 1967, Mallios, 1998).

Now, as we briefly alluded to in the introduction, the essential result from
(Raptis, 2000{\it a}), and the one that qualifies
finsheaves as sound approximations of the continuous
spacetime observables on $X$, is that an inverse system of finsheaves
has an inverse limit topological space that is homeomorphic to $S(X)$-the
sheaf of continuous functions on $X$, in the same way that in (Sorkin, 1991)
an inverse system of finitary poset substitutes of $X$ was seen to `converge'
to a space that is homeomorphic to the continuous topological manifold $X$ itself.

To define finsheaves of qausets we adopt from
(Raptis and Zapatrin, 2000) the association with every poset finitary
substitute $F_{n}$ of a bounded spacetime region $X$, of a Rota incidence
algebra $\omg(F_{n})$, as it was shown in the previous section.
As $F_{n}$ is a topological poset, its
associated $\omg_{n}$ is a topological incidence algebra (Raptis, 2000{\it
b}). As we noted in the previous section, to get the
qauset $\amg_{n}$ from $\omg_{n}$, we `causalize' and `quantize' it {\it
\`a la} Raptis (2000{\it b}). As a result of such a causalization, we write
$\vec{\rho}$ for the
generating relation of $\amg_{n}$'s (quantum) causal topology in the same
way that $\rho$ in the previous section was seen to be the generator
of $\omg$'s spatial Rota
topology. The significance of $\vec{\rho}$ is (quantum) causal, while of
$\rho$, only topological.

Finsheaves of qausets are then defined to be objects
$\ess_{n}:=\amg_{n}(\vec{F}_{n})$\footnote{Note again that we have put
an arrow over both
$F_{n}$ and $S_{n}$, since the partial order relation $\rightarrow$ is interpreted
causally rather
than topologically (Sorkin, 1995, Raptis, 2000{\it b,c}).}, whereby the
local homeomorphism between the base causal set $\vec{F}_{n}$ and
$\amg_{n}$ is now given, in complete analogy to the finsheaf $S_{n}(F_{n})$ of
topological posets in (Raptis, 2000{\it a}), as $p\imm q\Leftrightarrow
I_{p}\vec{\rho}I_{q},~(p,q\in\vec{F}_{n},~I_{p},I_{q}\in\vec{\cal{S}}
(\amg_{n}))$\footnote{See previous section for relevant definitions and note
that the primitive spectrum ${\cal{S}}$ of the qauset $\amg_{n}$ also carries an arrow
over it to remind one of its causal meaning.}. As
it was mentioned in the previous section, the Sorkin poset topology on the
topological $F_{n}$, obtained as the transitive closure
of the immediate contiguity relation $\imm$
between its vertices, is the same as the Rota topology of its
associated topological incidence
algebra $\omg_{n}$ generated by $\rho$\footnote{In other words, $\rho$ for
the Rota topology of $\omg_{n}$ is the transitive reduction of Sorkin's partial order
topological relation $\rightarrow$ in $F_{n}$.}, only now, these
relations have a directly causal/temporal rather than a topological/spatial significance
(Sorkin, 1995, Raptis, 2000{\it b}).

Also, in the same way that
$S_{n}(F_{n})$ was seen to be the finsheaf of continuous maps on the
$F_{n}$ obtained from $X$ with respect to its locally finite open cover
${\cal{U}}_{n}$ and generated by its (germs of) continuous sections (Raptis, 2000{\it a}),
we may similarly consider
${\bf G}_{n}:={\cal{L}}_{n}(\amg_{n})$ to be the finsheaf of local (quantum) causal
(auto)morphisms of $\amg_{n}$. We may call ${\bf G}_{n}$ `the
finitary spacetime transformation sheaf adjoint to $\ess_{n}$'\footnote{${\bf
G}_{n}$ is a group sheaf with carrier or representation or associated
sheaf that of qausets $\vec{S}_{n}$. The proper technical name for
${\bf G}_{n}$ is `principal sheaf with structure group
${\cal{L}}_{n}$' although, as we also mentioned in the introduction,
we use the latter denomination for the pair $(\vec{S}_{n},{\bf
G}_{n})$ (see below).}.
The (germs of) continuous sections of this sheaf are precisely the maps
that preserve the local (quantum) causal topology $\vec{\rho}$ of $\amg_{n}$ and by the
definition of the latter, they are the
$\amg_{n}$-homomorphisms `restricted'
to the primitive ideals $I_{p}$ and $I_{q}$ in them-the Gel'fand
`point-events' of the qauset $\amg_{n}$\footnote{This topological
interpretation of the primitive ideals of an incidence algebra $\omg$
associated with a finitary poset substitute $F$ in (Sorkin, 1991) as
`space points', comes from the Gel'fand `spatialization procedure' used in
(Zapatrin, 1998, Breslav {\it et al.}, 1999), whereby,
the point-vertices of the poset substitute $F$ of $X$
were corresponded to elements of the primitive spectrum
${\cal{S}}$ of its associated incidence algebra $\omg$ which, in turn, are
the kernels of the irreducible representations of $\omg(F)$
(see previous section). In our causal version $\amg$ of $\omg$, the primitive
spectrum of the former is denoted by $\vec{\cal{S}}$ and its points
({\it ie}, the primitive ideals of $\amg_{n}$) are
interpreted as `coarse spacetime events' ({\it ie}, they are equivalence
classes of $X$'s point-events relative to our pragmatic observations
${\cal{U}}_{n}$ of them of `limited power of resolution') (Raptis and Zapatrin,
2000, Raptis, 2000{\it a}).} which is the finitary
base space of the
finsheaf ${\bf G}_{n}$. The finsheaf ${\bf G}_{n}$ consists of the local causal
homeomorphisms $\vec{\lambda}_{n}$ of the local (quantum) causal topology
(generated by)
$\vec{\rho}$ of the qauset $\amg_{n}$ which, by the discussion in section 2,
constitutes the finitary version of the orthochronous Lorentz group
$L^{+}$\footnote{Hence ${\bf G}_{n}$ may be thought of as the finitary
substitute of the continuous Lie group-manifold $L^{+}$ which, due to
the (local) quantal character of the qausets in $\amg_{n}$, also inherits
some of the latter's `quantumness' in the sense that since qausets coherently
superpose with each other locally according to the FLSP of section 2, so will
their symmetry transformations. This is in accord with Finkelstein's
insight that if spacetime is to be regarded as being fundamentally a quantum system,
then so must be its structure symmetries (Finkelstein, 1996). See
also discussion in ($b^{'}$) of section 6.}.
Thus, the finsheaf $\ess_{n}$,
together with its adjoint ${\bf G}_{n}$ of its local symmetries,
constitute a principal ${\bf G}_{n}$-finsheaf of qausets and their finitary
local causal (and quantal) homeomorphisms\footnote{Which may be algebraically
represented by the group of incidence algebra homomorphisms as we mentioned
in section 2. If, as we
noted above, we allow for coherent quantum superpositions between these
local quantum causal symmetries as we allow for the causal connections
themselves in $\amg_{n}$, then presumably this group is a `quantum group' in a
sense akin to (Finkelstein, 1996). Interestingly enough, the Lie algebra
$sl(2,\com)$ has been shown (Selesnick, 1994) to result from the quantization of the classical binary
alternative ${bf 2}$-the local symmetry of the dyadic cell of the quantum causal
net in (Finkelstein, 1988). However, we are not going to
explore further the `quantum group' ({\it ie}, the quantal) character of the finitary local causal
symmetries in the group finsheaf ${\bf G}_{n}$.}. We may denote this principal
finsheaf either by
$\vec{\cem}_{n}:={\bf G}_{n}(\ess_{n})$, or more analytically by
$\vec{\cem}_{n}:=(\vec{F}_{n},\amg_{n},{\cal{L}}_{n})$\footnote{The symbol
`$\vec{\cem}_{n}$' for `${\bf G}_{n}(\ess_{n})$' will be explained shortly.},
with the corresponding local
homeomorphisms defining them as finsheaves (with a causal topological
interpretation) being denoted as
$\vec{s}_{n}:~(\vec{F}_{n},\imm)\rightarrow(\amg_{n},\vec{\rho})$ and
$\vec{\lambda}_{n}:~(\amg_{n},\vec{\rho})\rightarrow(\ess_{n},{\bf g}_{n})$;
where the reticular local
causal homeomorphism $\vec{\lambda}_{n}$ corresponds a $\vec{\rho}$-preserving
map to an element in the reticular Lie algebra ${\bf g}_{n}$ of the
structure group ${\bf G}_{n}=L^{+}_{n}$ of
the ${\bf G}_{n}$-finsheaf $\vec{\cem}_{n}$\footnote{That is, the reticular version of
the Lie algebra $\ell^{+}$
of the orthochronous Lorentz group $L^{+}=SO(1,3)^{\uparrow}$ whose
algebraic structure
is supposed to respect the `horizontal reticular causal topology' of
$\amg_{n}$
which is generated by $\vec{\rho}$-`the germ of the local quantum causal topology' of
the qauset stalks $\amg_{n}$ of ${\bf G}_{n}$'s associated finsheaf
$\vec{S}_{n}$ (Raptis, 2000{\it a}).}.

The main conjecture in this paper, briefly mentioned at the end of
(Raptis, 2000{\it a}) and in the introduction, and not to be analytically
proved here, is that an inverse
system ${\cal{K}}$ of the ${\bf G}_{n}$-finsheaves of qausets
$\vec{\cem}_{n}$ converges to the classical flat Minkowskian ${\bf G}$-sheaf
$(X\subset\cem ,\omg ,\ddd ,L^{+})$, where $X$ is a
bounded region in the smooth, flat Minkowski
manifold $\cem$, which serves as the base space for the sheaf
of smooth differential forms $\omg$ on it. This sheaf has as stalks over
$X$'s point-events
isomorphic copies of the $\integer$-graded module of Cartan exterior
differential forms
$\omg:=\omg^{0}\oplus\omg^{1}\oplus\omg^{2}\ldots$, $\ddd$ is the nilpotent
and flat K\"ahler-Cartan
connection on the sheaf\footnote{As we mentioned in the introduction,
$\ddd$ effects the following (sub)sheaf morphisms in the differential triad
$(X,\omg ,\ddd)$; $\ddd :~\omg^{i}\rightarrow\omg^{i+1}$
(Mallios, 1998).}, while $L^{+}$ is the structure group of the
sheaf consisting of the global orthochronous Lorentz transformations of
$\cem$\footnote{This description of the sheaf
$(\cem,\omg,\ddd ,L^{+})$ makes it the ${\bf G}$-sheaf theoretic analogue of a
${\bf G}$-bundle of exterior forms having as base space the flat
Minkowski differential
manifold $\cem$, as fibers modules of smooth Cartan forms
on $\cem$, as flat generalized differential ({\it ie}, connection) structure
the nilpotent
K\"ahler-Cartan differential $\ddd$, and as structure group the
orthochronous Lorentz group $L^{+}$. One may regard this sheaf as the
mathematical structure in which classical as well as quantum field theories
(excluding gravity) are formulated.}. Heuristic arguments that support
this conjecture are:

($a$) The topological ({\it ie}, $C^{0}$) structure of ($X$ of) $\cem$ as a
topological manifold arises as the limit space of an inverse system of
finitary incidence algebras $\omg_{n}(F_{n})$, now topologically interpreted, as
shown in (Sorkin, 1991, Raptis and Zapatrin, 2000). It can also be determined
from the causally interpreted incidence algebras $\amg_{n}(\vec{F}_{n})$ as the
quote from Bombelli {\it et al.} (1987) given in the introduction
suggests.

($b$) The differential ({\it ie}, $C^{\infty}$-smooth) structure of ($X$ in)
$\cem$ as a differential manifold supporting fibers of modules $\omg$
of Cartan's exterior forms, arises as the limit space of an inverse
system of finitary incidence algebras $\omg_{n}(F_{n})$, since the latter have
been seen to be discrete differential manifolds in the sense of Dimakis and
M\"uller-Hoissen (1999) (Raptis and Zapatrin, 2000, Zapatrin 2000).
In fact, as Dimakis {\it et al.}'s paper (1995) shows, the discrete differential
structure of such discrete differential manifolds also determines their
(finitary) topology\footnote{That is to say, ``{\it differentiability implies
continuity}"-the classical motto in mathematical analysis.}. The differential structure of $\cem$ can also be
determined from the causally interpreted incidence algebras $\amg_{n}(\vec{F}_{n})$
as the quote from Bombelli {\it et al.} (1987) given in the introduction
suggests.

($c$) The Minkowski metric $\eta_{\mu\nu}=diag(-1,+1,+1,+1)$ on $\cem$ is
determined by the causal incidence algebra $\amg_{n}$ associated with the
causal set $\vec{F}_{n}$ as the quote from (Bombelli {\it et al.}, 1987)
also suggests\footnote{As we mentioned in the introduction, the work of Robb
(1914) already shows that
causality as a partial order determines a Lorentzian metric up to its
determinant (spacetime volume-measure). See also (Bombelli and Meyer, 1989, Sorkin, 1990).}.
It must be emphasized however that in order to determine an
indefinite Lorentzian spacetime metric such as $\eta_{\mu\nu}$, the causally
interpreted
finitary incidence algebras must be used, not the topological ones. This is
because, as it was shown in (Dimakis and M\"uller-Hoissen, 1999), the
discrete metric $g$ that is naturally defined on a discrete differential
manifold such as the finitary topological incidence algebra of
(Raptis and Zapatrin, 2000), is positive definite
(Riemannian), rather than indefinite (pseudo-Riemannian, Lorentzian).
This is the `signature problem' alluded to in the introduction. The solution
of the `signature problem' by using causets instead of topological posets
justifies Finkelstein's (1988) and Sorkin's (1995) demand for a physical
causal or temporal topology instead of an unphysical spatial one, as we discussed in
the previous section.

($d$) The K\"ahler-Cartan differential operator $\ddd$ that defines
the differential structure of $\cem$ in ($b$) is a flat connection on
the differential triad sheaf $(\cem ,\omg ,\ddd)$ (Mallios, 1998),
as it is expected
to be for the flat Minkowski base space $\cem$. In (Dimakis and
M\"uller-Hoissen, 1999), a connection $\nabla$ and its
associated curvature $R:=-\nabla^{2}$ are defined, and compatibility conditions between
$\nabla$ and the definite metric $g$ are given that make the connection a
metric one\footnote{That is to say, a connection satisfying
$\nabla g=0$.}. However, since as it was mentioned in ($c$), $g$ is a
positive definite metric, $\nabla$ will not do, for we are looking for a
pseudo-Riemannian (Lorentzian) connection on our finsheaves of qausets.
Furthermore, as
it was also shown in (Dimakis and M\"uller-Hoissen, 1999), for the most general
(universal) discrete differential calculus on a discrete differential
manifold, $\nabla$ reduces to the flat (because nilpotent) K\"ahler-Cartan
differential $\ddd$, so that
there is no discrete (not even positive definite-Riemannian) `gravity' on it.
This is the `flatness problem' alluded to in the introduction. The flatness
problem will be tackled in the next section by a straightforward
localization or gauging of qausets in their finsheaves.

($e$) Finally, for the sheaf of global orthochronous Lorentz transformations
that we expect to arise as the group sheaf (Mallios, 1998) of (global)
symmetries of its adjoint\footnote{Or `associated'.} flat Minkowskian sheaf $(\cem ,\omg ,\ddd)$ from
an inverse system of
finsheaves ${\bf G}_{n}={\cal{L}}_{n}(\amg_{n})$ in the same way that
the flat differential triad
$(\cem ,\omg ,\ddd)$ arises from an inverse system of the
finsheaves $\ess_{n}=\amg_{n}(\vec{F}_{n})$, the work of Zeeman
(1964) provides significant clues. The key idea from (Zeeman, 1964) for our
finitary considerations here is that when causality is modeled after
a partial order between events in $\cem$, its causal automorphisms ({\it
ie}, one-to-one, order-preserving maps of $\cem$) constitute a group
${\bf G}$ isomorphic to the orthochronous Lorentz group $L^{+}$. Also, ${\bf
G}$ is, by definition, the group of homeomorphisms of $\cem$ regarded as a causal space
having for topology the causal Alexandrov (1956, 1967) one (Torretti, 1981).
It follows that the maps in the finsheaf ${\bf G}_{n}$, being by definition
local homeomorphisms of the qauset $\amg_{n}$, respect the
local (quantum) causal topology of $\amg_{n}$ which, in turn, effectively
corresponds to the
generating relation $\vec{\rho}$. These are the finitary (and quantal) analogues
of the causal automorphisms in (Zeeman, 1964), as we argued earlier. In fact, in the next section,
by a heuristic implementation of the FEP, FLP and FLRP given in section 2,
we will use these finitary causal
morphisms to define a finitary, quantal and causal gauge theoretic version of
Lorentzian gravity on the gauged $\vec{\cem}_{n}$\footnote{That is to say, the
principal finsheaf $\vec{\cem}_{n}$ will be supplied with a non-flat ${\bf g}_{n}$-valued
spin-Lorentzian connection ${\cal{A}}_{n}$ and its associated
curvature ${\cal{F}}_{n}$.}. For the
time being we note that the expected Minkowskian classical limit ${\bf G}$-sheaf
$(X\subset\cem ,\ddd ,\omg,L^{+})$, being flat, admits of global sections
(Mallios, 1998), a result which in physical parlance is known by the
following fact: `there is a global inertial coordinate patch (frame or gauge) covering
flat Minkowski space' (Torretti, 1981). However, in a curved spacetime $M$,
there are only local inertial frames (gauges) `covering' ({\it ie}, with origin) its
point-events by the CEP. These are independent (of each other) kinematical
frames (gauge possibilities) as we said in section 2
and this `kinematical independence' or `gauge freedom'\footnote{To be
explained in the next section.} motivates us here to define a non-flat connection
on ({\it ie}, to gauge) the flat ${\bf G}_{n}$-finsheaf $\vec{\cem}_{n}$.
Then, the resulting gauged, hence curved, finsheaf will not admit global
sections\footnote{Of particular interest is that it will not admit a global Lorentz-valued connection
section ${\cal{A}}_{n}$ of its $\amg^{1}_{n}$ sub-sheaf.} (Mallios, 1998).

We close this section by commenting on the operational significance of our
${\bf G}_{n}$-finsheaf model of quantum causality and its (global) causal
symmetries. If we take seriously the
conjecture above about the convergence of the system
${\cal{K}}:=\{\vec{\cem}_{n}\}$
to the classical flat Minkowskian ${\bf G}$-sheaf
$(X\subset\cem ,\omg ,\ddd ,L^{+})$ at maximum resolution of $X$ into
its point-events {\it \`a la} Sorkin (1991) and Raptis (2000{\it a}),
then sound operational meaning may be given to
qausets and their finitary symmetries in complete analogy to
the one given to topological poset substitutes $F_{n}$ of bounded regions $X$
of continuous spacetime manifolds $M$ in (Sorkin, 1991, 1995) and
their quantal algebraic relatives $\omg_{n}(F_{n})$ in
(Raptis and Zapatrin, 2000). Since the
$F_{n}$s were seen to converge to $X$, they were taken to be sound approximations of
its point-events, whereby a coarse determination of the locus of an event $x$
in $X$ is modeled by an open set about it. We emulate this semantic model
for the $F_{n}$s in the case of our qausets $\amg_{n}$ as
follows: we introduce a new `observable'\footnote{Actually, to be established as a dynamical variable
in the next section where we gauge $\vec{\cem}_{n}$.} for spacetime events called `causal
potential (or propensity) relative to our locally finite (coarse)
observations ${\cal{U}}_{n}$ of them', and
symbolized by $\pot$, so that the causal relation
$x\rightarrow y$ between two events in $\vec{F}_{n}$ can be read as `$x$ has
higher causal potential than $y$' ({\it ie}, formally: $\pot(x)>\pot(y))$\footnote{This formal labeling of events by $\pot$ is in complete analogy to the natural number $\N$-labeling of events {\it \`a la} Rideout and Sorkin (2000). There the sequential growth dynamics proposed for causets was seen to be independent from their $\N$-labeling thus in some sense `gauge independent' of an external (background) discrete $\N$-valued time gauge. In the next section we will see that similarly the reticular gauge connection ${\cal{D}}_{n}$ based on which the dynamical law for qausets will be formulated as an equation between sheaf morphisms, will be seen to be gauge ${\cal{U}}_{n}$-independent, thus also ${\cal{A}}_{n}$-covariant. See also (Raptis, 2000{\it e}) for more on this.}.
Thus, causality may be conceived as `causal potential difference between
events relative to our observations of them'\footnote{In a plausible `particle interpretation'
of our reticular scheme, whereby a network of causet (or qauset) connections is interpreted in the
manner of Dimakis {\it et al.} (1995) as the
reticular pattern of the dynamics of particles (or quanta) of
causality which may be called `causons' for obvious reason,
the causal connection $x\rightarrow y$ has the following rather natural physical interpretation in terms
of the causal potential $\pot$: `a causon descends from the event $x$ of higher causal potential to the
event $y$ of lower causal potential'. This is in literal analogy, for instance, with the motion of
an electron in an electromagnetic potential gradient, hence the
natural denomination of $\pot$ as `causal potential'.}.

This definition of $\pot$ applies in case the causal set $\vec{F}_{n}$
is the causally interpreted finitary poset substitute $F_{n}$ of a bounded
spacetime region $X$ as defined in (Raptis, 2000{\it b}). If $F_{n}$ derives from
the locally finite open cover ${\cal{U}}_{n}$ of $X$, $\pot$ in
$\vec{F}_{n}$ may be read as follows: the causal potential $\pot$ of an event
$x$ in $X$ relative to our observations ${\cal{U}}_{n}$ of
$X$\footnote{One can
equivalently call it `the causal potential of an event $x$ in $X$ at the
limit of resolution of $X$ corresponding to ${\cal{U}}_{n}$'
(Raptis, 2000{\it a}). The definition of $\pot$ as being `relative to our
coarse spacetime observations' is reflected by its index which is the same as
that of the locally finite open cover ${\cal{U}}_{n}$ of $X$-a finite $n$
signifying a pragmatic limited (finite), but at the same time coarse and perturbing,
power of resolution of $X$ into
its point-events.} corresponds to the `nerve' ${\cal{N}}$ covering $x$ relative to ${\cal{U}}_{n}$,
whereby ${\cal{N}}(x):=\{ U\in{\cal{U}}_{n}|~ x\in U\}$
(Raptis and Zapatrin, 2000)\footnote{As we also saw in the previous section,
in (Raptis and Zapatrin, 2000) nerves were seen to
be simplicial complexes and the topological discretization of manifolds based
on them is due to Alexandrov (1956, 1961). In (Raptis and Zapatrin, 2000),
the degree of a nerve of a covering is its cardinality, namely, the
number of the open sets in the covering that constitute it.}. Then, at the
level of resolution of the spacetime manifold-now regarded as a causal
space, corresponding to
$\vec{F}_{n}$, $x\rightarrow y$ ({\it ie}, `$x$ causes $y$') means
operationally that ${\cal{N}}(y)\subset {\cal{N}}(x)$ ({\it ie}, `every
(rough) observation of $y$ is a (coarse) observation of
$x$')\footnote{See (Breslav {\it et al.}, 1999) for a
similar operational semantics, but applied to the topological not to the
causal structure of spacetime like we do here.}; thence, $\pot(x)>\pot(y)$.
In terms of the definition of the smallest open
sets in ${\cal{U}}_{n}$ containing $x$ and $y$,
$\Lambda(x)$ and $\Lambda(y)$, given in section 3 and in (Sorkin, 1991,
Raptis, 2000{\it b}), that is to say,
$\Lambda(x):=\bigcap\{ U\in{\cal{U}}_{n}|~ x\in U\}\equiv
\bigcap {\cal{N}}(x)$, ${\cal{N}}(y)\subset{\cal{N}}(x)$ reads
$\Lambda(x)\subset\Lambda(y)$ with `$\subset$' standing for strict
set theoretic inclusion\footnote{Simplicially speaking, `$x$ is a face of $y$ with respect to
${\cal{U}}_{n}$'. See section 3 and (Breslav {\it et al.}, 1999, Raptis and Zapatrin, 2000).}.
This is precisely how the topological partial order $\rightarrow$ in
$F_{n}$ was
defined in (Sorkin, 1991), only in our $\vec{F}_{n}$ it is re-interpreted
causally (Raptis, 2000{\it b})\footnote{Note that event-vertices in the
causet $\vec{F}_{n}$ that are causally unrelated ({\it ie}, in some
sense `space-like') are covered by different
nerves in ${\cal{U}}_{n}$ of equal degree or cardinality (Raptis and Zapatrin, 2000).}. It must be
mentioned that such a conception of (quantum) causality as a
difference in cardinality (or degree) was first conceived in a different
mathematical model by Finkelstein (1969), while in (Breslav {\it et al.},
1999), and in a model similar to ours, the collection ${\cal{U}}_{n}$ of
open sets were assigned to
teams (or organizations) of `coarse observers' of spacetime topology and it is
explicitly mentioned that the relation $x\rightarrow y$ means that ``{\it
the event $x$ has been observed more times {\rm (by the team)} than the
event $y$}". However there, the `$\rightarrow$' obtained from Sorkin's
`equivalence algorithm' (section 3) is seen to still have its original
topological meaning and is not given a directly causal significance like in
our scheme\footnote{That is, the quantum observable or dynamical variable in their theory is topology, not local causality.}.

>From the definition of $\pot$ above, it follows that the generator of local
({\it ie}, contiguous\footnote{See section 2.}) causal
potential differences between events in $\vec{F}_{n}$ corresponds to the relation of
immediate causality $\imm$ linking events, say $x$ and $y$, such that
$\Delta\pot(x,y):=\pot(x)-\pot(y)=1$. That is to say, we may symbolize this `contiguous'
causal potential difference-the `germ' of the quantum causal potential, by
$\impot$. If we pass to the qauset $\amg_{n}$ associated with
$\vec{F}_{n}$, or equivalently, to the finsheaf $\ess_{n}$ of
qausets, the aforementioned generator of
causal potential differences assumes a completely algebraic expression as
$\vec{\rho}$. Again, we recall from section 3 that
$I_{p}\vec{\rho}I_{q}\Leftrightarrow I_{p}I_{q}(\not= I_{q}I_{p})
\stackrel{\not=}{\subset} I_{p}\cap I_{q}$ generates
the quantum causal Rota topology of $\ess_{n}$ by relating primitive
ideals $I_{p}$ and $I_{q}$ in $\vec{{\cal{S}}}(\amg_{n})$
($p,q\in\vec{F}_{n}$) if and only if $p\rightarrow q$ and
$\not\!\exists r\in\vec{F}_{n}:~ p\rightarrow r\rightarrow q$ ({\it ie}, iff
$p\imm q$ in $\vec{F}_{n}$) (Breslav {\it et al.}, 1999, Raptis, 2000{\it b}).

Here, in the algebraic setting of qausets,
the generator of quantum causality ({\it ie}, the germ $\impot$ of the quantum
causal potential $\pot$) relative to our finitary spacetime
observations in ${\cal{U}}_{n}$,
$\vec{\rho}$, has the following operational and quantal {\it \`a la}
Heisenberg (because non-commutative algebraic) meaning that reads from
its very algebraic
definition: point-events in $\amg_{n}$, which correspond to primitive ideals
in $\vec{\cal{S}}(\amg_{n})$\footnote{Recall from (Breslav {\it et al.},
1999, Raptis and Zapatrin, 2000, Raptis, 2000{\it b}) and section 3
the definition of the primitive ideals in the corresponding
quantum topological $\omg_{n}(F_{n})$:
$I_{p}:=\spn\{|q><r|:~|q><r|\not=|p><p|\}$; where $|q><r|:=(q\rightarrow
r)\in F_{n}$. Parenthetically, it is rather interesting to observe in this definition of the
primitive ideals (points) in the quantal topological incidence
algebras $\omg_{n}(F_{n})$ that the elements (ket-bras) that constitute
them are quantal acts of determination of what in the classical limit space
will emerge as `momentum (covector) states' and serial concatenations thereof
({\it ie}, `spacetime time-like paths'); see physical interpretation of the $\omg^{i}$s ($i\geq 1$) in
(Raptis and Zapatrin, 2000). By this very definition of the $I_{p}$s in
${\cal{S}}(\omg_{n}(F_{n}))$, we see that the operations of determination of pure
quantum spacetime states (events) in $\omg_{n}$, namely, the elements of
$\omg^{0}$ (the $|p><p|$s in the definition of the $I_{p}$s above;
Raptis and Zapatrin, 2000), are excluded from them. So, operations of
determination of what classically ({\it ie}, at the non-pragmatic decoherence limit of
infinite refinement of the spacetime continuum into its point-events)
appear as momentum states tangent to
spacetime `position states' (point-events) are `incompatible' or
`complementary' in Bohr's sense with ({\it ie}, they
exclude) quantum acts of localization of the latter. This observation
shows that some kind of
quantum uncertainty is built into our Rota algebraic scheme {\it ab initio}
thus it further justifies the physical interpretation of the limit of infinite
localization of spacetime events as Bohr's correspondence principle (Raptis
and Zapatrin, 2000). The quantum character of the non-commutative topology generated by the local (and dynamical) quantum causality $\vec{\rho}$ is analytically studied in (Raptis, 2000{\it e}).}, have a product ideal that is strictly included
in their intersection ideal, with the `directedness' (asymmetry) of their immediate
quantum causal connection, say `from-$p$-to-$q$' ($p\imm q$), being reflected in the
non-commutativity of their corresponding ideals in $\amg_{n}$
({\it ie}, $I_{p}I_{q}\not=
I_{q}I_{p}$)\footnote{This is a first indication of a fundamental
non-commutativity of (acts of localization of) `points' ({\it ie}, `coarse
spacetime events') underlying quantum causal topology in a model like ours (where
`points' are represented by primitive ideals in the primitive spectra
$\vec{\cal{S}}$ of the incidence algebras $\amg$ involved). In a coming paper
(Raptis, 2000{\it e}), the incidence algebras modeling qausets
here, as well as their localizations, are studied in the light of scheme theory (Hartshorne, 1983, Shafarevich, 1994) and a non-commutative dynamical local quantum causal topology for (at least the kinematics of) Lorentzian quantum gravity is defined based on such non-abelian schematic algebra localizations in much the same way to how Non-Commutative Algebraic Geometry was defined in (Van Oystaeyen and Verschoren, 1981) based on non-abelian Polynomial Identity (PI) ring localizations-it being understood that Rota algebras can be regarded as PI rings (Freddy Van Oystaeyen in private communication). It must be a fruitful project to compare the resulting `non-commutative topology for curved quantum causality' in (Raptis, 2000{\it e}) with the one defined and studied in (Van Oystaeyen, 2000{\it a}). The second author (IR)
wishes to thank Freddy Van Oystaeyen for motivating such a study in a crucial
private communication and in two research seminars; see (Van Oystaeyen, 2000{\it b}).
Ultimately, the deep connection for physics is anticipated to be one between such
a non-commutative conception of the local quantum causal topology of spacetime and the
fundamental micro-local quantum time asymmetry expected of `the true quantum gravity' (Penrose,
1987). Again, such a fundamental time asymmetry in a curved finitistic quantum causal topological space similar to ours has already been anticipated by Finkelstein (1988).}. This operational description of quantum
causality in $\amg_{n}$ relative to our coarse observations of events in a bounded region of
spacetime-now interpreted as a causal space, follows from the operational
description of causality in the causet $\vec{F}_{n}$ from which it
derives via the causal potential `observable' $\pot$ defined above.

All in all, (quantum) causality is operationally defined and interpreted
as a `power relationship' between spacetime events relative to our coarse
observations (or approximate operations of local determination) of
them, namely, if events $x$ and $y$ are coarsely determined by ${\cal{N}}(x)$
and ${\cal{N}}(y)$ with respect to ${\cal{U}}_{n}$, and ${\cal{N}}(y)\subset
{\cal{N}}(x)$, then `$x$ causes $y$'. The attractive feature of such a
definition and interpretation of causality is that, by making it relative to
${\cal{U}}_{n}$, we render it `frame- or observation-dependent'\footnote{One
may think of
the open $U$s in ${\cal{U}}_{n}$ as some sort of `rough coordinate patches'
or `coarse frames' or even as `fat gauges' (Mallios, 1998) `covering' or coarsely
measuring (approximately localizing) the point-events in $X$.}, ultimately,
relativistic\footnote{Recall that the causal potential $\pot$ of events is
defined relative to our coarse observations ${\cal{U}}_{n}$ of them, so that,
as we will see in the next section, its localization (gauging) and
relativization will effectively amount to establishing a local transformation
theory for it that respects its dynamics (due to a finitary sort of
Lorentzian quantum gravity), in the sense that this dynamics becomes
independent of the level of resolution corresponding to our observations
${\cal{U}}_{n}$ of spacetime into its events, or equivalently, it becomes
independent of the local
gauges (frames) ${\cal{U}}_{n}$ that one lays out to chart the spacetime
events and measure, albeit coarsely, physical attributes such as the gravitational field
`located there' (Mallios, 1998). This will be then the
transcription of the fundamental principle of GR, which
requires that the laws of physics are invariant under the diffeomorphism
group of the smooth spacetime manifold ${\rm Diff}(M)$ (the principle of General
Covariance), in a sheaf theoretic model for a curved finitary quantum
causal space: `the laws of physics are equations expressed in terms of
sheaf morphisms'-the main sheaf morphism being the connection ${\cal{D}}$
(Mallios, 1998). We will return to this principle in section 5 where we
define ${\cal{D}}_{n}$ as a finsheaf morphism in our scheme and
further discuss its quantum physical implications in section 6.}.

In the same way, one can give operational meaning to the finitary local
(quantum) causal
automorphisms of the $\amg_{n}$s in $\ess_{n}$ mentioned
above. They represent
finitary operations of `approximation'\footnote{The inverted commas for the
word `approximation' were explained in the introduction.} of the local
symmetries of quantum causality as encoded in the finsheaf $\ess_{n}$
and they too are organized in the finsheaf
${\bf G}_{n}$. The operational interpretation of the elements of
${\bf G}_{n}$ as coarse reticular (and quantal) replacements of the continuous local orthochronous Lorentz Lie
symmetries of the smooth gravitational spacetime of GR will
become transparent in the next section when we gauge the flat Minkowskian
${\bf G}_{n}$-finsheaf $\vec{\cem}_{n}$ by providing a non-flat
${\bf g}_{n}$-valued connection $1$-form ${\cal{A}}_{n}$ on it.

\section*{\normalsize\bf 5. GAUGING QUANTUM MINKOWSKI SPACE: NON-FLAT
CONNECTION ON ${\bf\ess_{n}}$}

\indent\sloppy{The reader was prepared in the previous sections for the present
one where we will attempt to curve the flat and quantal
${\bf G}_{n}$-finsheaf $\vec{\cem}_{n}:={\bf G}_{n}(\ess_{n})$ by gauging or
localizing it. As it was mentioned earlier, this procedure is
tantamount to defining a reticular non-flat spin-Lorentzian
connection ${\cal{A}}_{n}$\footnote{The reader should note the index
$n$ given to the connection ${\cal{A}}$ that is the same as the one given
to the causet $\vec{F}_{n}$, its associated qauset $\amg_{n}$ and
the latter's local quantum causal symmetries ${\cal{L}}_{n}$.
Properly viewed, the connection ${\cal{A}}$ on the ${\bf G}_{n}$-finsheaf
$\vec{\cem}_{n}:=(\vec{F}_{n},\amg_{n},\ddd_{n},{\cal{L}}_{n})$ in focus
inherits the latter's `finite degree of resolution' $n$ of the region $X$ of the
curved spacetime manifold $M$ by our
coarse and dynamically perturbing observations ${\cal{U}}_{n}$ of its events, their causal ties and
the symmetries of the latter (Raptis, 2000{\it a}). The
reader should notice that the index $n$ is also given to the reticular K\"ahler-Cartan
differential $\ddd$ in $\vec{\cem}_{n}$ just to remind one of its discrete
character {\it \`a la} Dimakis and M\"uller-Hoissen (1999).}
that takes values in the Lie algebra ${\bf g}_{n}$ of the group finsheaf ${\bf G}_{n}$
adjoint to $\ess_{n}$ consisting of the latter's local quantum causal
symmetries-the finitary and quantal
substitute of the continuous orthochronous Lorentz Lie group manifold
$L^{+}$ which, in
turn, is the structure group of (global symmetries of) the flat
${\bf G}_{n}$-finsheaf
$\vec{\cem}_{n}$\footnote{As we contended in the
last two sections and briefly alluded to in the introduction, this is
the finitary, causal and quantal substratum underlying the flat principal
Minkowskian sheaf
$(X\subset\cem,\omg ,\ddd, L^{+})$ of modules of Cartan differential forms
over the bounded region $X$ of the Minkowski manifold $\cem$, equipped with
the flat K\"ahler-Cartan exterior differential (connection) $\ddd$ and having as
continuous structure group $L^{+}$ that of (global) orthochronous Lorentz
symmetries of $\cem$. This `classical' sheaf, being flat, admits of global
sections of
its adjoint structure group sheaf $L^{+}$ over $\cem$ (Mallios, 1998).}.
The resulting curved ${\bf G}_{n}$-finsheaf
$\vec{\cal{P}}_{n}:=(\vec{F}_{n},\amg_{n},{\cal{L}}_{n},{\cal{D}}_{n}:=
\ddd_{n}+{\cal{A}}_{n})$ may be regarded as a finitary, causal and quantal replacement of
the classical structure ${\cal{P}}$ on which GR is formulated as a gauge theory of a
spin-Lorentzian connection $1$-form
${\cal{A}}$. This model ${\cal{P}}$ of the curved classical spacetime structure of GR,
as it was noted in the introduction, is a
principal fiber bundle of modules $\omg$ of Cartan
differential forms, over a region $X$ of a $C^{\infty}$-smooth
Lorentzian spacetime manifold $M$, with structure group
$L^{+}:=SO(1,3)^{\uparrow}$ and
a non-flat $so(1,3)^{\uparrow}$-valued gravitational gauge connection
${\cal{A}}$ on it. As it was also briefly noted in the introduction,
this classical model may be equivalently thought of as a rigorous mathematical
formalization of the spinorial formulation of GR due to
Bergmann (1957) when, fiber-wise ({\it ie}, locally), one corresponds the
real Minkowski spacetime vectors dual to the Minkowskian co-vectors
$\omg^{1}$ in the Cartan
bundle, to $H(2,\com)$-the space of Hermitian bispinors (or Hermitian
biquaternions) (Finkelstein, 1996) in the complex completely antisymmetric
tensor bundle
$T^{*}\simeq\omg^{1}_{\com}\simeq S\otimes\tilde{S}\equiv(\tilde{S}^{*}\otimes
S^{*})^{*}$\footnote{Where `$*$' denotes `dual space' and
$T\simeq(\tilde{S}^{*}\otimes S^{*})$ is fiber-wise isomorphic to complex
Minkowski space $\com^{4}$ (Selesnick, 1991, 1994, 1998).}
over the $2$-dimensional Grassmannian subspaces
$Gr_{2}(\com^{4})$ of complexified Minkowski space $\com^{4}$\footnote{The
spinorial formulation of GR, as well as the self-dual
gravity of Ashtekar (Ashtekar, 1986, Baez and Muniain, 1994), inevitably involve
complexified spacetime. Imposing reality conditions or recovering
real spacetime ($\real^{4}$) from a complex model (Finkelstein, 1988, Selesnick, 1994, Baez
and Muniain, 1994) is certainly not an easy business or without
physical significance. Likewise, being incidence algebras, our qausets may be
regarded as vector spaces over the field $\com$, so, if anything, the finitary
and quantal version of $\cem$ as a causal space presented in sections 3
and 4 is surely complex in case we use $\com$ for linear coefficients
(amplitudes) over
which qausets superpose coherently in the $\amg_{n}$ stalks of $\vec{\cem}_{n}$
(Raptis, 2000{\it b}). However, we are not going
to present here the transition from complex to real spacetime and
gravity. On the other hand, we must be careful not to infuse {\it ab initio} the real or the complex
number continua into our inherently reticular scenario, for then we will be begging the question
`discrete before continuous ?'.
Which amplitude (c-coefficient) structure one must use for qausets and other algebraic structures that have been suggested to model
quantum spacetime and gravity is still an open problem
(Chris Isham in private correspondence).} (Manin, 1988, Selesnick, 1991). This correspondence is the usual
one: $x_{\mu}\rightarrow x_{\mu}\sigma_{\mu}=:x_{a\dot{b}}\equiv
s\otimes s^{\sim};~(s\in S\simeq\com^{2}, s^{\sim}\in \tilde{S}
\simeq\dot{\com}^{2})$\footnote{$S$ is the space of spinors transforming under the fundamental irreducible
representation of $SL(2,\com)$, while $\tilde{S}$ consists of the conjugate
spinors that transform as vectors under the conjugate (and inequivalent to
the fundamental) irreducible representation of $SL(2,\com)$. See
(Selesnick, 1991, 1994, 1998) for notation and relevant definitions.};
with $\sigma_{\mu}$ the usual tetrad of Pauli's
$2\times 2$ spin-matrices.

Then, again fiber-wise, the correspondence
between the structure groups of the
Bergmann and the Cartan principal fiber bundles is the usual projective one
given by the $2$-to-$1$ map: $\rho :~SL(2,\com)\rightarrow
SO(1,3)^{\uparrow}=:L^{+}$, which reflects the fact that $SL(2,\com)$ is the
double cover of $SO(1,3)^{\uparrow}$. However, again as it was briefly
mentioned in the introduction, locally in the group fiber
({\it ie}, Lie algebra-wise), the two structure groups are isomorphic, since
$so(1,3)^{\uparrow}\simeq sl(2,\com)$\footnote{Hence our calling ${\cal{P}}$
`the Cartan-Bergmann ${\bf G}$-bundle' and ${\cal{A}}$ on it `the spin-Lorentzian
connection'.}.

Finally, as it was also alluded to in the introduction, our holding that
the curved ${\bf G}_{n}$-finsheaf $\vec{\cal{P}}_{n}$
is a finitary, causal and quantal replacement of the classical Cartan-Bergmann
${\bf G}$-bundle ${\cal{P}}=(X,\omg,L^{+},{\cal{D}}:=\ddd+{\cal{A}})$ model of GR, basically rests on the
idea that an inverse system of the former curved finsheaves
of qausets yields, at the
operationally ideal limit of finest resolution or localization of $X$ into
its point-events, their causal ties and the local symmetries thereof, which
limit, in turn, may be interpreted as Bohr's correspondence principle
(Raptis and Zapatrin, 2000), the latter as a classical gravitational
spacetime structure (Raptis, 2000{\it a}).

Thus, we consider a bounded region $X$ of a curved smooth spacetime manifold $M$.
We assume that gravity is represented by a non-flat $sl(2,\com)$-valued
connection $1$-form $\cal{A}$ on the curved Cartan-Bergmann ${\bf G}$-bundle
${\cal{P}}=(X,\omg,L^{+},{\cal{D}}=\ddd+{\cal{A}})$.
First we discuss a mathematical technicality that our finsheaf theoretic
model should meet in order to be able to define a (non-flat)
connection ${\cal{D}}_{n}$\footnote{Note that until now we used
the gauge potential ${\cal{A}}$ for the mathematical concept of connection
${\cal{D}}$, when, in fact, $\cal{A}$ is just the part of ${\cal{D}}=\ddd+{\cal{A}}$ that
makes it non-trivial ({\it ie}, non-flat) (Mallios, 1998). This is the
physicist's `abuse' of the concept of connection, presumably due to his rather
`utilitarian' or at least `practical'
attitude towards mathematics, namely, that he is interested on the part of
${\cal{D}}$
that is responsible for curvature (which can be physically interpreted as
the gauge potential of a physical force). In fact, the substitution
$\ddd\rightarrow {\cal{D}}=\ddd+{\cal{A}}$ is coined `gauging' in the physics
jargon, when $\ddd$ is from a mathematical point of view a perfectly legitimate connection;
albeit, a trivial ({\it ie}, flat) one (Mallios, 1998). The same `abuse' of ${\cal{D}}$ is
encountered
in (Baez and Muniain, 1994; see chapter 5) where only the gauge potential $\cal{A}$ is
coined `connection'. Here, we too adopt a physicist's approach and by
`gauging our flat Minkowskian principal finsheaf $\vec{\cem}_{n}$'
essentially we mean `adjoining a non-zero connection term ${\cal{A}}_{n}$
to its flat differential $\ddd_{n}$'. This
asymphony between the mathematician's and the physicist's conception of the
notion of
connection aside, one should always keep in mind that ${\cal{D}}$ is a
generalized differential operator, with
its non-zero part $\cal{A}$ generalizing or extending the usual differential
operator $\ddd$.} on the (flat) finsheaf $\vec{\cem}_{n}$. Two sufficient
conditions for the existence of a connection ${\cal{D}}$ on an algebra
or vector sheaf or bundle over a manifold $M$, regarded as a topological
space, are that $M$ is paracompact and Hausdorff\footnote{We recall that a
topological space $M$ is said to be
paracompact if every open cover of it admits a locally finite refinement.
Also, $M$ is said to be Hausdorff or $T_{2}$, when it satisfies the second axiom of
separation of point-set topology which holds that every pair of points of $M$ have non-intersecting
(disjoint) open neighborhoods about them.} (Mallios, 1998). It is expected that,
since our finsheaves of qausets are finitary (and quantal) replacements of an
at least $T_{1}$\footnote{We recall that a topological space
$X$ is said to be $T_{1}$ if for every pair of points $x$ and $y$ in it there exist
open neighborhoods $O_{x}$ and $O_{y}$ containing them such that $x\not\in O_{y}$ and
$y\not\in O_{x}$.} and relatively compact\footnote{As it was also noted earlier,
a topological space $X$
is said to be relatively compact, or bounded in the sense of (Sorkin, 1991),
when its closure is compact.} topological space $X$ (Sorkin, 1991, Raptis,
2000{\it a}), if we relax $T_{2}$ to $T_{1}$ and paracompactness to relative
compactness, we are still able to define a connection ${\cal{D}}_{n}$ on a vector or
algebra sheaf over it such as our $\vec{\cem}_{n}$ (Raptis, 2000{\it a}).
So ${\cal{D}}_{n}$ exists ({\it ie}, is `defineable') on $\vec{\cem}_{n}$.
In fact, we know that since $\ddd_{n}$ is already defined on the
qauset stalks of $\vec{\cem}_{n}$ {\it \`a la} Dimakis and M\"uller-Hoissen
(1999)\footnote{See also (Zapatrin, 2000).} and it is
seen to effect sub-sheaf morphisms
$\ddd_{n}:~\amg^{i}_{n}\rightarrow\amg^{i+1}_{n}$ there (Mallios, 1998);
albeit, it is a flat connection (Mallios, 1998, Dimakis and
M\"uller-Hoissen, 1999). In turn, this ${\cal{D}}_{n}=\ddd_{n}$ on
the finitary, causal and quantal Minkowskian finsheaf $\vec{\cem}_{n}$
means that ${\cal{A}}_{n}=0$ throughout $\vec{\cem}_{n}$, so that by
our physical terminology the latter is an ungauged (flat)
finsheaf\footnote{As we also mentioned in the previous section, the
important point to retain from the discussion above is that in a sheaf theoretic
context like ours the role of connection ${\cal{D}}$ is as a sheaf morphism (Mallios,
1998). We will come back to it shortly when we formulate a
finsheaf theoretic version of the Principle of General Covariance of GR in
the gauged $\vec{\cem}_{n}$.}.

To curve the flat finsheaf $\vec{\cem}_{n}$ by adjoining to its flat
connection $\ddd_{n}$ a non-zero term ${\cal{A}}_{n}$,
we immitate in our finitary context how the curved smooth spacetime manifold
$M$ of GR may be thought of as the result of localizing or
gauging the flat Minkowski space $\cem$ of SR.
Locally, ({\it ie}, event-wise), one
raises\footnote{That is to say, as if it is a `vertical structure'.} an
isomorphic copy of $\cem$ over each spacetime event $x\in X\subset M$, thus
implementing the CEP of section 2. Hence formally, $\cem$ acquires an
event-index $x$ ($\forall x\in X$), ${\cem}_{x}$, and may be regarded as
some kind of `fiber space over $x$'\footnote{As we mentioned in section 2,
${\cem}_{x}$ may be physically interpreted as
a local kinematical (possibility) space for GR implementing
the kinematical CEP in the
sense that GR's main dynamical variable, $g_{\mu\nu}$, can
be locally reduced or `gauged away', by passing to a locally inertial frame,
to the flat $\eta_{\mu\nu}$ of the $\cem$ of SR.}. In view of the
differential ({\it ie}, $C^{\infty}$-smooth)
character of $M$, which, in turn, may be thought of as implementing the
CLP discussed in section 2 (Einstein, 1924),
${\cem}_{x}$ is geometrically interpreted as `the space
tangent to $M$ at $x$', so that collectively,
$TM:=\bigcup_{x\in M}{\cem}_{x}$ is the locally Minkowskian
tangent bundle of $M$ (G\"ockeler and Sch\"ucker, 1990) having for fibers
${\cem}_{x}$ ({\it ie}, local isomorphs of flat Minkowski space).

Then, the term `gauging' effectively corresponds to regarding these local
isomorphs of flat Minkowski space as `independent kinematical worlds', in the
sense that two vectors $v$ and ${v}^{'}$ living in the vector spaces
${\cem}_{x}$ and ${\cem}_{{x^{'}}}$, respectively, are `incomparable', in
that one is not allowed to form linear combinations thereof\footnote{For
instance, one is
not supposed to be able to compute their difference ${v}^{'}-v$ which
is the crucial operation for defining the differential operator $\ddd$ in general.}.
Alternatively, one may describe this in a geometrical way by saying that
in a gauged space,
such as the vector bundle $TM$ (G\"ockeler and Sch\"ucker, 1990), there is
no natural relation of distant parallelism between its fibers. A `rule' that
enables one to compare vectors at different fiber spaces, thus it
establishes some kind of relation of distant parallelism or, algebraically
speaking, `distant linear
combineability' in $TM$, is provided by the concept of connection
${\cal{D}}$ (Mallios, 1998). The geometrical interpretation of ${\cal{D}}$,
and one which shows an apparent dependence of this concept on the background
geometric spacetime manifold $M$\footnote{And we say `apparent', because,
as we will see shortly, in our scheme ${\cal{D}}_{n}:=\ddd_{n}+{\cal{A}}_{n}$
does not depend essentially on the geometric base spacetime, since it derives locally from the very
algebraic structure of the stalks of
the finsheaf of qausets ({\it ie}, from the
structure of the quantally and causally interpreted incidence
algebras). This is the main lesson we have learned from the Abstract
Differential Geometry theory developed in (Mallios, 1998), namely,
that ${\cal{D}}$, the main object with which one can actually do Differential
Geometry, is of an algebraic ({\it ie}, analytic) nature and does not depend
on any sort of `ambient geometric space'. For instance, the two global
(topological) conditions
for the existence of ${\cal{D}}$ on $M$ mentioned above, namely, that
the latter is a paracompact and Hausdorff topological space, are
sufficient, but by no means necessary. Such an independence is welcome
from the point of view of both classical and quantum gravity where the
spacetime manifold, regarded as an inert geometrical background base space,
has shown to us its pathological, `unphysical
nature' in the form of singularities and the non-renormalizable infinities
that plague the field theories defined on it.} is as a parallel transporter
of vectors along smooth curves
in $M$ joining $x$ with ${x^{'}}$. Then, curvature ${\cal{F}}$, in
the classical model
for spacetime corresponding to the differential manifold $M$, is
geometrically conceived as the anholonomy of ${\cal{D}}$ when the latter parallely
transports vectors along closed smooth curves (loops) in $M$-certainly a
non-local conception of the action of ${\cal{D}}$\footnote{In contradistinction
to this classical geometric
conception of curvature, in the sense that it depends on the existence of
spacetime loops in $M$ and that it is the effect of the action of ${\cal{D}}$ as
the parallel transporter of
geometric entities (smooth tensor fields) along them, we will be able to give
shortly a purely local sort of curvature ${\cal{F}}_{n}$ stalk-wise in our gauged
finsheaves of qausets.}.

The second problem that we face is one of physical semantics: we want to interpret
the non-flat part ${\cal{A}}_{n}$ of ${\cal{D}}_{n}$ in a finitary causal
way. In the classical curved spacetime model ${\cal{P}}$, $\cal{A}$, apart from its usual
interpretation as the gravitational gauge potential, may be physically
interpreted as the smooth dynamically variable (field of the) local causal
connections between the events of the $C^{\infty}$-smooth spacetime region
$X$\footnote{See section 2.}.
Since the fibers of the curved $TM$ (or the Minkowskian covectors in the
$\omg^{1}$ sub-bundle of ${\cal{P}}$) are local isomorphs of flat Minkowski
space, the action of the spin-Lorentzian gravitational connection
$1$-form $\cal{A}$ on Minkowski vectors living in $TM$'s fibers, besides its
geometrical interpretation as `parallel translation' above,
may be alternatively interpreted in a causal way as follows: point-wise
on the curve along which the vectors are transported, the transitive,
inertial Minkowskian causality $\rightarrow$
is preserved\footnote{One may conceive in this sense the standard requirement
in GR that `the connection is compatible with the metric
tensor field $g_{\mu\nu}$', or that `${\cal{D}}$ is a metric connection'
({\it ie}, formally, that ${\cal{D}}g=0$). This geometrical requirement for
`zero spacetime distortion' seems to be a convenience coming from the classical
continuum model $M$ of spacetime ({\it ie}, in pseudo-Riemannian geometry on
$M$, the metric $g_{\mu\nu}$ determines a unique metric connection
${\cal{D}}$-the usual Levi-Civita one $\Gamma$ (Torretti, 1981)), but it may not
hold in the quantum deep which assumes no pre-existent $M$,
let alone a $g_{\mu\nu}$ on it (Finkelstein, 1996). Like Finkelstein however,
only for convenience we assume that at every finite level $n$ of resolution of
spacetime ${\cal{A}}_{n}$ is metric-a convenience reflecting the fact that
in our finsheaf theoretical model the connection is naturally expected to respect
the local (quantum) Minkowskian causal topology $\vec{\rho}$ of the
qauset stalks (see sections 2-4, the discussion in the next paragraph and
in section 6).}. Equivalently put, if $x$ is a point in the curve and
$v(x)\in\cem_{x}$ is the value of a vector field $v$ at
$x$\footnote{Technically, a vector field is a cross-section of the vector
bundle $TM$ (G\"ockeler and Sch\"ucker, 1990).}, the `coupling'
${\cal{A}}(x)[v(x)]$\footnote{${\cal{A}}$ may be regarded as
a matrix ${\cal{A}}_{ij}$ of sections
of $1$-forms in $\omg^{1}$ each taking values in the Lie algebra ${\bf g}$
of the structure group fiber (stalk) ${\bf G}_{x}$ of the ${\bf G}$-bundle
(sheaf) in focus. This is essentially Cartan's definition
of connection (Von Westenholz, 1981, G\"ockeler and Sch\"ucker, 1990,
Mallios, 1998). In our case, it suffices to define $\cal{A}$ as a
(matrix of) section(s) of
the ${\bf G}$-sheaf $(X,\omg^{1},L^{+})$ taking values in the Lie algebra of
the structure group ${\bf G}=L^{+}=SO(1,3)^{\uparrow}$ of orthochronous Lorentz transformations
(Baez and Muniain, 1994).} may be thought of as a local
Lorentz transformation (`infinitesimal spacetime rotation') of $v(x)$;
hence, by definition, it preserves
the local causal structure of the curved $TM$, namely, the Minkowski
lightcone based at (or with origin) $x$ in $\cem_{x}$\footnote{One could also
say `the Minkowski lightcone opening up in the fiber $\cem_{x}$ soldered at
$x$'.}. In this sense, one may
equivalently interpret the gravitational gauge connection $\cal{A}$ as the
dynamical field of local causality, as we noted in section 2. We adopt
this physical interpretation for our finitary ${\cal{A}}_{n}$.

Now we come to the crucial point of the present paper that was briefly
mentioned at the end of section 2 and in the footnote above concerning metric
connections. From a causal
perspective, like the one we have adopted here, a sheaf may be the
`natural' mathematical structure to model such (dynamically variable) local causal
connections such as $\cal{A}$, because of the following rather heuristic
argument which in a sense motivated us to study finsheaves of qausets
in the first place: by definition, a sheaf is a local homeomorphism
(Bredon, 1967, Mallios, 1998, Raptis, 2000{\it a}), so that when one is
interested in the (dynamically varying) causal topology of spacetime like we are in the present
paper where our ${\bf G}_{n}$-finsheaf of qausets is supposed to be the
`quantum discretization' (Raptis and Zapatrin, 2000) of
the local causal topology ({\it ie}, the causal connections between events)
and its local symmetries of a bounded region
$X$ of a curved smooth spacetime manifold $M$, a sheaf preserves precisely
the (germs of the) local causal topology of the base space\footnote{See
previous section and (Raptis, 2000{\it a}).}.
But, in our case, the latter are precisely the immediate causality (contiguity)
relations $\imm$ in
the causal set $\vec{F}_{n}$ that are mapped by the sheaf (regarded as a
local homeomorphism $\vec{s}$\footnote{See previous section.}) to the
germ-relations $\vec{\rho}$ of the quantum causal Rota topology of
the qausets $\amg_{n}$, thus defining the finsheaf $\ess_{n}$
of qausets over the causal set\footnote{Again, see previous section for relevant
definitions.}. Also, the adjoint sheaf ${\cal{L}}_{n}$, it too regarded
as a local homeomorphism
$\vec{\lambda}$\footnote{Also see previous section.}, preserves the
generator ({\it ie}, the generating relation or `local germ') $\vec{\rho}$ of the quantum causal
topology of $\amg_{n}$, thus it consists of local, finitary causal and
quantal versions of the orthochronous Lorentz group
$L^{+}=SO(1,3)^{\uparrow}$\footnote{These are the
finitary and quantal analogues of infinitesimal causal automorphisms of
Minkowski space. As we have said before, the latter, by Zeeman's work (1964), may be identified with
the spin-Lorentz Lie algebra $\ell^{+}=so(1,3)^{\uparrow}\simeq sl(2,\com)$
of $L^{+}$.}. Altogether, a local $\amg^{1}_{n}$-section\footnote{The reader
should note the arrow over the sub-sheaf space $\omg^{1}_{n}$ of discrete
$1$-forms in $\vec{\cem}_{n}$ which again shows its causal interpretation, as
well as its `finite resolution index' $n$.}
of the ${\bf G_{n}}$-finsheaf ${\cal{L}}_{n}(\ess_{n})$ associates, via the composition
$\vec{\lambda}_{n}\circ\vec{s}_{n}$ of the two
local homeomorphisms defining the finitary sheaf $\ess_{n}$ and its
adjoint ${\cal{L}}_{n}$, with a
contiguous causal arrow $x\imm y$ in the causet $\vec{F}_{n}$ a reticular Lorentz
local (infinitesimal) transformation in $[\ell^{+}_{n}]_{x}$\footnote{Note
again the index $n$ we have
given to the local quantum causal symmetries of $\amg_{n}$. In fact, one
should write $I_{x}$ instead of $x$ for the index of the stalk of the
finsheaf ${\bf G}_{n}(\ess_{n})$,
since the point-event in the qauset $\amg_{n}$ corresponding to
the event $x$ of the causal set $\vec{F}_{n}$ is the primitive ideal
$I_{x}$ of the quantum causal incidence algebra $\amg_{n}$ associated with
the causet $\vec{F}_{n}$ (see sections 3 and 4).} which, in turn,
may be thought of as `rotating' the quantal Minkowskian vectors in the stalk
$[\amg_{n}]_{[x]}$\footnote{See (Raptis, 2000{\it a}) for notation and
definition of stalks of finsheaves $S_{n}(F_{n})$.} of
$\amg_{n}(\vec{F}_{n})$ over the event
$x$. Now we see how natural it is to define a finitary spin-Lorentzian connection
${\cal{A}}_{n}$ as a local $\amg^{1}_{n}$-section of the ${\bf G}_{n}$-finsheaf
$(\vec{F}_{n},\amg_{n},\ddd_{n},{\cal{L}}_{n})=:\vec{\cem}_{n}$.

However, as we said earlier, since the latter
is flat, it admits global sections (Mallios,
1998). Flatness means that ${\cal{A}}_{n}\equiv 0$ throughout
$\vec{\cem}_{n}$ (Mallios, 1998), or equivalently, that `the
connection is identically equal to the trivial constant zero global
$\amg^{1}_{n}$-section of the ${\bf G}_{n}$-finsheaf $\vec{\cem}_{n}$'.
In our finitary causal context, we attribute this to the
constancy ({\it ie}, the non-dynamical character) and the transitivity of
the inertial Minkowskian causal connection $\rightarrow$ in $\vec{F}_{n}$,
a property that is
certainly non-local (Finkelstein, 1988, Raptis, 2000{\it b})\footnote{See the
CEP and its finitary formulation FEP in section 2.}. In fact, the
`unphysicality' of a chrystalline-rigid causality relation modeled after a
transitive, and due to this, global partial order, is already implicitly
noted by Zeeman (1964)\footnote{Remark 3 in (Zeeman, 1964). Parenthetically,
we may recall from (Zeeman, 1964) that if $\cem$ is Minkowski space and for any
two events $x,y\in\cem$ $x<y$ means that $y-x$ is a future-directed
timelike vector ({\it ie}, $\| y-x\|:=
(y-x)^{\mu}\eta_{\mu\nu}(y-x)^{\nu}<0$;
$\eta_{\mu\nu}=diag(-1,1,1,1)$ and $y_{0}>x_{0}$), then the group ${\bf G}$ of
all one-to-one maps $f$ of $\cem$
to itself that preserve `$<$' ({\it ie}, the `causality group'
consisting of the causal automorphisms of $\cem$), is isomorphic to the
inhomogeneous conformal orthochronous Lorentz group
({\it ie}, the orthochronous Lorentz group $L^{+}$,
together with translations, dilatations and space-reversals, but with
spacetime volume-preserving maps being excluded). The latter reflects the fact
that causality, as a partial order, determines the Minkowski metric up
to its determinant (spacetime volume form) (Robb, 1914). Note also that
in this globally flat Minkowski case Zeeman can form the difference
$y-x$ of Minkowski vectors-a `distant comparability' that is not allowed
in the curved case (see discussion above).}:

\begin{quotation}
{\footnotesize The condition for $f$ to be a causal automorphism is a global
condition, but is equivalent (by an elementary compactness argument using
the transitivity of $<$) to the following local condition: given $x\in\cem$,
then there is a neighborhood $N$ of $x$ such that
$y<z\Leftrightarrow fy<fz,~\forall y,z\in N$. Intuitively this means we
need only think of the principle of causality acting in our laboratories
for a few seconds, rather than between distant galaxies forever, and
still we are able to deduce the Lorentz group...}
\end{quotation}

\noindent and explicitly by Finkelstein (1988) who also emphasizes the need
for a dynamical local causality:

\begin{quotation}
{\footnotesize ...the causal relation $x{\bf C}y$\footnote{Where the ${\bf C}$ in
Finkelstein's paper is the same as our constant transitive partial order causal
relation $\rightarrow$.} is not local, but may hold for events as far apart
as the birth and death of the universe. Since we have commited ourselves
to local variables, we abandon ${\bf C}$ for a local causal relation ${\bf
c}$... We localize the causal relation ${\bf C}$\footnote{We add `${\bf C}$'
at this point of the excerpt for emphasis.} by taking as basic dynamical variable
a relation $x{\bf c}y$ expressing {\it immediate} causal
priority\footnote{Where the ${\bf c}$ in Finkelstein's paper is the same as
the transitive reduction $\imm$ of the partial order causality relation
$\rightarrow$ of the causal set $\vec{F}_{n}$ (Raptis, 2000{\it b})-the germ
of the causal topology in the finsheaf $\ess_{n}$ of qausets
which, by the local homeomorphism $\vec{s}$ defining it, corresponds to the
germ $\vec{\rho}$ of the quantum causal
topology of the qausets $[\amg_{n}]_{x}$-the stalks over the coarse
point-events of $\vec{F}_{n}$.}.}
\end{quotation}

Indeed, like Finkelstein, we regard the germ relation $\imm$ of the local causal
topology of $\vec{F}_{n}$, or its finsheaf $\vec{s}$-image $\vec{\rho}$ of
$\amg_{n}$, as being dynamically variable. This is achieved by localizing or
gauging the finsheaf $\ess_{n}$ and its
adjoint ${\cal{L}}_{n}$\footnote{Of course, when one localizes or
gauges qausets, it follows that their local quantum causal
symmetries are gauged as well.} which, in turn, corresponds
to implementing a non-zero (non-flat) dynamically variable ${\bf g}_{n}$-valued
gauge connection ${\cal{A}}_{n}$ realized as a local
$\amg^{1}_{n}$-section of the ${\bf G}_{n}$-finsheaf
$\vec{\cal{P}}_{n}=(\vec{F}_{n},\amg_{n}, {\cal{L}}_{n},{\cal{D}}_{n})$. Thus,
${\cal{A}}_{n}$ effectively represents a finitary gravitational dynamics of
qausets.

Since ${\cal{A}}_{n}$ represents the dynamics of the germ
$\vec{\rho}$ of the quantum causal topology of the qauset-stalks of
$\vec{\cal{P}}_{n}$, it is defined locally\footnote{That is to say,
stalk-wise in the sheaf.}, thus purely algebraically\footnote{Since the
stalks are quantally and
causally interpreted incidence Rota algebras with the generator $\vec{\rho}$
of quantum causal topology being defined entirely algebraically with respect
to the stalks' spectra $\vec{\cal{S}}$ as we saw in sections 3 and 4.}. The
detailed algebraic argument that leads to the expression for ${\cal{A}}_{n}$
in terms of $\vec{\rho}$ is left for (Raptis, 2000{\it f}). For the present
paper it suffices to give the usual gauge theoretic expression for the curvature associated
with ${\cal{D}}_{n}$: ${\cal{F}}_{n}:={\cal{D}}_{n}^{2}={\cal{D}}_{n}\wedge{\cal{D}}_{n}=
[{\cal{D}}_{n},{\cal{D}}_{n}]$\footnote{Where `$\wedge$' denotes Cartan's exterior product
and `$[.,.]$'  `commutator'. It follows that ${\cal{F}}_{n}$ is a $\ell_{n}^{+}$-valued section of
the $\amg_{n}^{2}$ sub-sheaf of $\vec{\cal{P}}_{n}$.}
(G\"ockeler and Sch\"ucker, 1990, Baez and Muniain, 1994, Mallios, 1998, Dimakis and
M\"uller-Hoissen, 1999) and note that it is defined entirely locally-algebraically
stalk-wise in the sheaf without reference to an ambient geometric base
space. As we argued earlier this is quite welcome from the point of view
of `quantum gravity'\footnote{See also Einstein's quotation at the end of
($d$) in the next section.}. ${\cal{F}}_{n}$ may be physically interpreted rather freely
in our scheme as a finitary, causal and quantal expression of Lorentzian gravity.

Now that we have mathematically defined and physically interpreted
${\cal{A}}_{n}$ (and its curvature ${\cal{F}}_{n}$) on $\vec{\cal{P}}_{n}$,
we give an alternative physical
interpretation for it more in line with the operational interpretation of
finsheaves in (Raptis, 2000{\it a}), whereby, the latter were regarded
as `approximations of the continuous spacetime observables'. So again, let
$X$ be a bounded region in a curved smooth spacetime manifold $M$ on
which ${\cal{A}}$ lives (in
${\cal{P}}$ ). As in (Sorkin, 1991) the open sets $U$ in the locally
finite open cover ${\cal{U}}_{n}$ of $X$ were physically interpreted as
`coarse acts of
localization (local determination) of the continuous topology carried by
$X$'s point-events' which, in the finsheaf $S_{n}(F_{n})$ of (Raptis,
2000{\it a}), translates to
`coarse acts of local determination of the continuous ({\it ie},
$C^{0}$-topological) spacetime
observables'\footnote{Which by definition preserve the local Euclidean
manifold topology
of $X$.}, so similarly we lay out
${\cal{U}}_{n}$ to chart roughly the causal topology and the causal symmetries
of the bounded spacetime region $X$ in the gravitational spacetime
$M$\footnote{Following the terminology in Mallios (1998), we call the
elements $U$ of ${\cal{U}}_{n}$ `coarse (or fat) local
gauges', since they stand for rough acts of measurement of the local ({\it
ie}, the point-event) structure of $X$.}. Then, we organize our observations
of the dynamics of local quantum causality into the curved ${\bf G}_{n}$-finsheaf
of qausets $\vec{\cal{P}}_{n}$ as described above. In $\vec{\cal{P}}_{n}$,
we perceive as `gravitational gauge potentials ${\cal{A}}$' the ${\bf
g}_{n}$-valued $\amg^{1}_{n}$-sections ${\cal{A}}_{n}$. Now, in the manner
that we described the {\it aufbau} of $\vec{\cal{P}}_{n}$ in sections 2-5,
it is straightforward
to interpret ${\cal{A}}_{n}$ as `equivalence classes of gravitational gauge
potentials' relative to our coarse and dynamically perturbing causal topological observations
${\cal{U}}_{n}$\footnote{The reader must notice here that $X$ `physically
exists' as a background space to the extent that it lends itself to us as a
topological substrate on which we may lay (perform, apply, localize) our finitary
topological gauges ${\cal{U}}_{n}$-the locally finite pragmatic observations
of `it'. In
this sense it is a `surrogate space' which avails itself to us for performing
our quantum discretizations of `it' which, in turn, recover `it' as a local
({\it ie}, classical smooth) structure at the operationally implausible
(ideal),
classical limit of observations of infinite power (or energy) of resolution (Raptis and Zapatrin,
2000, Raptis, 2000{\it a}). See also remarks in (c) of the next section.}.
Equivalently, following verbatim the physical interpretation of
finsheaves in (Raptis, 2000{\it a}), ${\cal{A}}_{n}([x])$ stands for a
collection of gravitational gauge potentials that are `indistinguishable' at
the finite level $n$ of resolution of $X$ into its point-events\footnote{Thus, we
tacitly assume that the spacetime events are not only surrogate carriers of $X$'s
physical topology (Sorkin, 1991, Raptis, 2000{\it a}), but also of its other
physically observable fields (attributes)-the gravitational field being the
one in focus here.}. `Indistinguishability' may be physically interpreted
here in a dynamical way as follows: the gravitational field is not
perceived as varying between any two events in the same equivalence class
$[x]$\footnote{This interpretation is consistent with our FEP of section 2
which, in effect, held that in a finsheaf over a causet $\vec{F}_{n}$
obtained by ${\cal{U}}_{n}$ as described in sections 3 and 4, all the
frames in the `fat (coarse) stalks' $\amg_{n}([x])$ over $[x]$ are inertial
relative to
each other.}. Furthermore, and here is the operational weight that
the sheaf theoretic scheme of ours carries, it is our coarse operations of
determination of the dynamical local (quantum) causal topology, which are
organized into $\vec{\cal{P}}_{n}$, that are effectively encoded in
${\cal{A}}_{n}$, so that, `by the end of the day', it is not the
point-events of $X$ {\it per se} that carry information about the dynamics
of (quantum) causality; rather, it is our own dynamically perturbing observations
of them that create `it'\footnote{See the active operational interpretation of dynamical
local quantum
causality at the end of section 4. See also (c) in the next section.}. Then, the
General Relativistic character of
our sheaf theoretic scheme may be summarized in the following: the finitary
gravitational connection ${\cal{D}}_{n}$ on $\vec{\cal{P}}_{n}$ is a sheaf
morphism (Mallios, 1998), which means that the dynamics of local quantum
causality is ${\cal{U}}_{n}$-independent\footnote{For a short discussion of
this apparently paradoxical situation, namely, that our own coarse local
observations ${\cal{U}}_{n}$ create the dynamical local quantum causality whose
dynamics is subsequently expressed in a ${\cal{U}}_{n}$-invariant ({\it ie},
gauge independent) way, see (c) in the next section.}. This is the
(fin)sheaf theoretic version of the principle of General Covariance of GR
with a strong local-quantal flavor\footnote{Since ${\cal{D}}_{n}$ respects
the linear quantum kinematical structure ({\it ie}, the coherent quantum
superpositions of quasets) stalk-wise in $\vec{\cal{P}}_{n}$. See also
($b^{'}$) in the next section.}.

\section*{\normalsize\bf 6. DISCUSSION OF THE SOUNDNESS OF
${\bf\vec{\cal{P}}_{n}}$ AND OTHER RELEVANT ISSUES}

\indent\sloppy In this last section we present four arguments that support that
$\vec{\cal{P}}_{n}$ is a sound model of a finitary, causal and quantal
version of Lorentzian gravity.

\indent ($a$) The FEP of section 2 is satisfied in $\vec{\cal{P}}_{n}$,
since the latter's stalks $\amg_{n}$ are local isomorphs of a locally
finite, causal and quantal version of Minkowski space (sections 3 and 4).

The FLRP of section 2 holds in $\vec{\cal{P}}_{n}$, since the latter's group ${\bf
G}_{n}$-stalks are finitary, causal and quantal versions of the orthochronous Lorentz structure group $L^{+}$ of local causal symmetries of GR.

The FLP of section 2 is satisfied in $\vec{\cal{P}}_{n}$, since the latter's qauset stalks
are sound models of local quantum causality (section 3 and (Raptis, 2000{\it
b})).

The FLSP of section 2 holds in $\vec{\cal{P}}_{n}$, since the qausets
residing in its stalks coherently superpose with each other (sections 3, 4
and (Raptis, 2000{\it b})).

In section 2 we posited that a sound mathematical model of a finitary curved
quantum causal space should meet structurally these four `physical axioms'.
Indeed, the structure of $\vec{\cal{P}}_{n}$ does meet them.

($b$) $\vec{\cal{P}}_{n}$ is an algebraic model for finitary, local and dynamically
variable quantum causality which inherits its operational meaning from
the pragmatic and quantal interpretation given to quantum topological
incidence algebras in (Raptis and Zapatrin, 2000), hence also to their causal
relatives in (Raptis, 2000{\it b}). Moreover, from (Raptis and Zapatrin, 2000)
it inherits its essentially alocal character, while, together with the physical
interpretation of finsheaves (of algebras) in (Raptis, 2000{\it a}), it
manifests its essential non-commitment to spacetime as a background
geometrical $C^{\infty}$-smooth manifold.

($c$) The local structure of classical
gravitational spacetime, namely, the event and the space of Minkowskian
directions tangent to it, arise only at the operationally ideal limit of
infinite localization\footnote{That is to say, at the operationally ideal
situation of employment of an infinite power (or energy) to resolve spacetime into its
point-events. As we explained in section 3, this is theoretically
implausible too due to the fundamental conflict of the principles of
Equivalence and Uncertainty on which gravity and the quantum are founded at
energies ({\it ie}, microscopic powers of resolution) higher than
$\hbar t_{P}^{-1}$.} of an inverse system of $\vec{\cal{P}}_{n}$s (Raptis,
2000{\it a}). The latter limit, yielding the classical gravitational sheaf ${\cal{P}}$
in a manner analogous to how the sheaf $S(X)$ of continuous functions on
a topological spacetime manifold arises at infinite refinement of topological
finsheaves $S_{n}$ similar to our causal $\vec{\cal{P}}_{n}$s (Raptis,
2000{\it a}), may be physically interpreted as Bohr's correspondence principle (Raptis
and Zapatrin, 2000). This further supports the quantal character of
$\vec{\cal{P}}_{n}$. All in all, putting together the physical
interpretations of the theoretical schemes proposed in (Raptis and Zapatrin,
2000), (Raptis, 2000{\it b}) and (Raptis, 2000{\it a}) that are amalgamated
into our model $\vec{\cal{P}}_{n}$ for the dynamics of finitary quantum
causality as described in sections 3-5, we may summarize the
physical interpretation of $\vec{\cal{P}}_{n}$ to the following:
it respresents alocal, discrete, causal and quantal operations of determination of
the dynamics of causality and its symmetries in a bounded region of a
curved smooth spacetime manifold with the latter not existing in a
physically significant
sense\footnote{Not being `physically real', so to speak.}, but only viewed as
providing a surrogate scaffolding on which we base ({\it
ie}, locally solder\footnote{That is to say, we restrict our experimental activity
in laboratories of finite spatio-temporal extent. This is the operational
analogue of Zeeman's
remark in the quotation from (Zeeman, 1964) given in the previous section
that ``{\it we need only think of the principle of causality acting in our
laboratories for a few seconds, rather than between distant galaxies
forever}" although the time scale of observations of quantum causality is
expected to be of
the order of Planck time ($\approx 10^{-42}s$), not of few seconds.}) our own operations of
observing `it', which are then suitably organized into algebra finsheaves. This collective physical interpretation of $\vec{\cal{P}}_{n}$ is well in accord with
the general philosophy of quantum theory holding that inert, background,
geometrical `state spaces' and their structures, such as spacetime and its
causal structure, `dissolve away', so that what remains and is
of physical significance, the `physically real' so to speak, is (the
algebraic mechanism\footnote{Or `structure'.} of) our own actions of
observing `it' (Finkelstein, 1996).

In section 4 we stretched even further this
`observer-dependent physical reality' essence of quantum theory to an
`observation-created physical causality' with the introduction of the
`quantum causal potential relative to our coarse observations' observable which was
subsequently seen to be the dynamically variable entity represented by the
finitary connection ${\cal{A}}_{n}$ on $\vec{\cal{P}}_{n}$ (section 5) only
to find that a (fin)sheaf theoretic version of the principle of General Covariance
of GR holds in our model, namely, that dynamics is formulated
in terms of equations involving the connection ${\cal{D}}_{n}$ which is the
main finsheaf morphism in $\vec{\cal{P}}_{n}$ (Mallios, 1998). Thus,
our mathematical
expressions of
`physical laws' are not observation-dependent\footnote{That is to say, physical laws are
${\cal{U}}_{n}$-gauge independent or invariant under our coarse and dynamically
perturbing local
measurements of `spacetime' (Mallios, 1998).}. This points to the following
seemingly paradoxical interpretation of our scheme: the observer acts as a
`law-maker' when she observes\footnote{By establishing causal connections
between the events that she observes.} and as a `law-seeker' when she
communicates her observations\footnote{That is to say, when she `objectifies'
her actions of determination of `it' to others by organizing the coarse causal
nexus she has perceived in `it' to structures (sheaves) so that the dynamics
${\cal{D}}_{n}$ of this causal nexus (topology) is independent of her
`subjective' coarse measurements of `it all' in
${\cal{U}}_{n}$.}. There is no conflict, for as Finkelstein
(1996) notes:

\begin{quotation}

{\footnotesize ...Since we and our medium are actually a quantum entity, the
goal of knowing the dynamical law completely seems to be a typically
ontic\footnote{In (Finkelstein, 1996) `ontic' refers to the classical
existential ideal due to Plato which holds that `objects exist independently
of our own modes of perceiving them'. In physics, this Platonic ideal
translates to: `physical objects exist independently of our own actions of
determination of their physical properties'.} one. This completeness must
prove as counterphysical as the others we have already encountered.
Law-seekers are in some part law-makers as well. Just as we influence the
laws of geometry slightly by moving masses, we may influence any law of
dynamics at least slightly by our own actions...}

\end{quotation}

After all, such an apparently conflicting duality may be necessary for a
genuine synthesis of the quantum with relativity (Finkelstein, 1996)-a
synthesis which appears to be at the heart of the problem of
`quantum gravity' {\it per se}.

($d$) Causet theory (Bombelli {\it et al.}, 1987, Sorkin, 1990) addresses the problem of
`quantum gravity' in locally finite, causal and, to some extent, quantal
terms\footnote{For instance, a quantum dynamics for causets is sought after a covariant
path integral
or `sum over causet histories' scenario (Sorkin, 1990).} from a non-operational
point of view (Sorkin, 1995). Finkelstein's Quantum Net Dynamics (1988,
1989, 1991) and its subsequent generalization, Quantum Relativity Theory (1996), address
the same problem
in almost the same terms, but from an `entirely operational'\footnote{That
is to say, `pragmatic'.}
point of view. Our finsheaf theoretic model for finitary Lorentzian quantum
gravity brings together Finkelstein's and Sorkin's approaches under a `purely
algebraic roof' and to some
extent vindicates their fundamental insight that the problem
of `quantum gravity' may be solved or, at least, be better understood, if it
is formulated as `the dynamics of an atomistic local quantum causal
topology'. At least, it certainly goes some way to vindicate Einstein's hunch: ``{\it Perhaps
the success of the Heisenberg method points to a
purely algebraic method of description of nature, that is to the elimination
of continuous functions from physics}" (Einstein, 1936), and it
complies with his more general and imperative intuition (Einstein, 1956) that:

\begin{quotation}

{\footnotesize One can give good reasons why reality cannot at all be
represented by a continuous field. From the quantum phenomena it
appears to follow with certainty that a finite system of finite energy
can be completely described by a finite set of numbers (quantum
numbers). This does not seem to be in accordance with a continuum
theory, and must lead to an attempt to find a purely algebraic theory
for the description of reality. But nobody knows how to obtain the
basis of such a theory.}

\end{quotation}

We conclude the present paper by briefly discussing six physico-mathematical
issues that derive from our finsheaf model of finitary Lorentzian quantum
gravity:

($a^{'}$) The first issue is about localization. The duality between
topological Rota incidence algebras and poset finitary substitutes has been
established (Raptis and Zapatrin, 2000). Similarly, the physical duality
between spacetime observables and spacetime states (point-events) was
discussed in (Raptis, 2000{\it a}). Categorically speaking, this duality
corresponds to a
contravariant functor from the category of poset-finitary
substitutes/poset-morphisms to the category of topological incidence
algebras/algebra-homomorphisms (Raptis and Zapatrin, 2000, Zapatrin, 2000)\footnote{This functorial equivalence between topological posets and their incidence algebras is used in a most fruitful way in (Raptis, 2000{\it e}) to define a `quantum topos' of scheme theoretic localizations of qausets (see $f^{'}$ below).}.
Localizations in the first category are
represented by inverse limits as in (Sorkin, 1991)\footnote{Such inverse
limits yield the finest or `smallest' open sets containing the spacetime
point-events in $X$ (Sorkin, 1991).}, while in the second, by
direct limits\footnote{Such direct limits yield the finest fiber spaces of
continuous functions over $X$'s point events $x$-the stalks of the sheaf in focus
consisting of germs of continuous functions at $x$ for every $x$ in $X$
(Raptis, 2000{\it a}).}. Inverse and direct limits are formal (categorical)
processes of topological localization dual (opposite)
to each other (Dodson, 1988). In the present paper, where we also dealt with
operations of localization of the main causal-topological observable
${\cal{A}}_{n}$ in a curved finitary causal space, the finsheaf approach with
its `coarse direct limit localizations' as in (Raptis, 2000{\it a}) seems
rather natural to adopt for qauset-localizations (to `coarse or fat stalks') and for
the coarse gravitational gauge potential ${\cal{A}}_{n}$ that twists them `there'.

($b^{'}$) Related to our `localization' comments above, is that
in a finitary curved quantum causal space similar to ours, Finkelstein's
causal net (1988), gravity may be thought of as arising from breaking quantum
coherence in the net-a phenomenon which gives rise to vectors in flat Minkowski space
(Selesnick, 1994). Since the latter constitute the local structure of a
curved spacetime by the CEP, it seems that the very acts of
event-localization are
responsible for this lifting of `quantum coherence', thus rendering the FLSP
of section 2 ineffective. Hence, something like (classical) spin-Lorentzian
gravity results when one attempts to localize or gauge Finkelstein's quantum
spacetime net by breaking its quantum coherence (Selesnick, 1994). Similarly
in our scheme, ${\cal{A}}_{n}$, which is the result of localizing or gauging
quantal Minkowskian qausets in the finsheaf $\vec{\cal{P}}$, may be thought
of as somehow breaking the local (kinematical) quantum superpositions of qausets
stalk-wise, by establishing incoherent `distant linear combineability of
qausets'\footnote{That is to say, it allows for superpositions of qausets
that live in different stalks in $\vec{\cal{P}}$. See section 5.}. On the
other hand, the metric ${\cal{A}}_{n}$, by taking values in the local Rota algebra
homomorphisms of the qauset-stalks, respects their local linear
superpositions stalk-wise. There is no contradiction if we assume that at
finite and still quantal level of resolution of $\vec{\cal{P}}_{n}$ the
gravitational connection still preserves local coherent quantum suprpositions
of qausets\footnote{In a sense, by locally preserving the quantum coherence of
qausets ({\it ie}, stalk-wise), ${\cal{A}}_{n}$ qualifies for a quantal sort
of gravitational action. It follows
that, since the qauset `paths' (Raptis and Zapatrin, 2000, Raptis, 2000{\it b})
coherently interfere, so does the
gravitational connection ${\cal{A}}_{n}$ that is defined on them. Thus, some
kind of reticular ({\it ie}, `innately or inherently regularized') path integral
quantization of gravity, by
considering for instance a local spin-Lorentz covariant path integral quantization
scenario over the space of
qauset connections ${\cal{A}}_{n}$ on the principal finsheaves thereof, is
already implicit in our scheme. However, the ambitious project to explicate
this and put it on a rigorous footing must be left for a future investigation.}
and only at the ideal classical limit of infinite point-event
localization coherence is broken and classical gravity ${\cal{A}}$ on
${\cal{P}}$ emerges as in (Selesnick, 1994)\footnote{It is interesting to
mention at this point that Selesnick ascribes the emergence of classical
gravity on Finkelstein's quantum net by localizing or gauging the $SL(2,\com)$
symmetries of the quantum relativistic binary alternative to ``{\it what a macroscopic
observer having sufficiently limited powers of spacetime resolution
perceives as this gauging}". Our finitary model
$\vec{\cal{P}}_{n}$ for Lorentzian gravity also initially derives from
such pragmatic coarse observations of the topological structure of spacetime
as in (Sorkin, 1991, Raptis, {\it a}), but in the manner of (Raptis, 2000{\it
b}) it also acquires its essentially causal and quantal interpretation, thus
unlike Selesnick's account, its gravity should be placed at the quantum and
not the classical side of Heisenberg's {\it schnitt}. Only at the ideal limit
of infinite spacetime localization does classical gravity emerge (on
${\cal{P}}$).}.

Furthermore, Finkelstein (1988) has entertained the idea that such
quantum coherence-breaking localization processes are in fact physical quantum
condensations of the net\footnote{Hence the epithet `superconducting' given
to such quantum causal nets (Finkelstein, 1988).} and classical spacetime
is thought of as arising from such `phase transitions' that the net
undergoes. It would certainly be worthwhile to try and relate this theory
to our curved qausets and their decoherence to classical gravitational spacetime.

($c^{'}$) The third issue is about spacetime dimensionality. As we
mentioned in the introduction, we did not consider questions of
dimensionality in the present paper. However, since dimensionality is a
topological invariant and our finsheaf scheme is supposed to be an account
of (at least the kinematics of) a dynamical local quantum causal topology, it should somehow address the problem
of the dimensionality of spacetime and its dynamical fluctuations at quantum
scales. Bombelli {\it et al.} (1987) and Sorkin (1990), for instance, entertain
the idea that spacetime dimensionality
is a statistical variable varying with the power of `fractally' refining or
resolving causets. On the other hand, as it was also mentioned at the end of
(Raptis and Zapatrin, 2000), albeit in a quantum topological not a causal context
like ours, our qausets should not be thought of as being of statistical
nature so that the
classical spacetime dimensionality, as well as the other classical spacetime
properties such as the differential and local Lorentz-causal structures
arise at some `thermodynamic limit' as if they are `statistical average
attributes of spacetime'. Rather, since classical spacetime is expected to
arise at the classical decoherence limit from curved finsheaves of qausets
in a way similar to
Finkelstein's net condensation scenario, we too anticipate that spacetime
dimensionality is a long range order parameter of the condensate
(Finkelstein, 1988, 1996)\footnote{Thus, from a qauset point of view, spacetime dimensionality
is a quantum variable, but from a causet point of view it is a classical random variable, and there
is a significant difference between these two kinds of dynamical variables (David Finkelstein in
private correspondence). We may attribute this distinction to the possibility of coherent quantum
superpositions between qausets and the absence of such a possibility for causets (Raptis, 2000{\it b}).}.

Furthermore, Finkelstein (1996) anticipates that not only the spacetime
dimensionality but also
that the spacetime metric $g_{\mu\nu}$, the basic dynamical variable in GR,
is such a long range parameter of a quantum condensate, so that
the requirement for the connection to be metric\footnote{That is to say,
that there is no spacetime distortion (Finkelstein, 1996).} is only for our
theoretical convenience and not a truly fundamental aspect of Nature. Our
theoretical model $\vec{\cal{P}}_{n}$ of finitary Lorentzian quantum gravity
also invites the assumption of a non-distorting connection ${\cal{A}}_{n}$,
thus the latter is `model-dependent'\footnote{That is to say, `theoretically convenient'. See previous section.} and may prove to be physically not fundamental.

($d^{'}$) The fourth issue is about `space-likeness versus time-likeness'.
At the end of (Sorkin, 1991), in a comparison between topological posets
and causets, it is explicitly mentioned that the former, say $P$, may be used to
represent the coarse spatial topology of a `thickened'\footnote{We
equivalently use the epithet `fat' for `coarse' approximations (see above).} space-like Cauchy
hypersurface in a globally hyperbolic spacetime manifold $M$ which
approximates a fundamental causet $C$\footnote{Notice here that in the same
way that we criticized the term `approximations' when applied to describe
qausets in the introduction, Sorkin also thinks of the classical spacetime
manifold $M$ as an approximation to causets, not the other way around.}.
Our qausets on the other hand are quantum causal replacements of the continuous spacetime $4$-manifold and
are essentially of time-like nature; they are not
topological approximations-proper to spatial $3$-submanifolds of $M$. The principle of
classical spacetime causality, which is modeled by assuming that $M$ is
globally hyperbolic, may also be a long range (global) spacetime property
arising from the decoherence of an ensemble of fundamental alocal qauset substrata.

($e^{'}$) The fifth issue we would like to address here concerns gravity
as a Yang-Mills type of gauge theory. In the Einstein-Cartan theory
(G\"ockeler and Sch\"ucker, 1990) or in Ashtekar's self-dual formulation (Ashtekar,
1986, Baez and Muniain, 1994) the gravitational spin-Lorentzian gauge
potential ${\cal{A}}$\footnote{In the Einstein-Cartan theory it is customary
to write $\omega$ for the spin connection instead of ${\cal{A}}$. $\omega$ is
also what Selesnick obtains from gauging Finkelstein's net as briefly
alluded to above (Selesnick, 1994). Ashtekar uses only the self-dual
part ${\cal{A}}^{+}$ of the gravitational gauge potential.} tells only half of the gravitational
story in that there is also the
(complex) frame field $e^{I}_{a}$\footnote{Also known as `co-moving tetrad'
or `{\it vierbein}'.} gravitational variable. `$I$' is an internal (spin)
index, while `$a$' an external spacetime index. Put together into a
Palatini-like action functional and varied independently of each other the two
gravitational variables ${\cal{A}}$ and $e^{I}_{a}$ yield two sets of
equations: variation with respect to ${\cal{A}}$ yields an equation between
it and the usual Christoffel metric connection $\Gamma$ of GR on $M$ (in
${\cal{P}}$) which
establishes the equivalence between the original metric formulation of GR in
terms of the Levi-Civita $\Gamma$ and the gauge theoretic one in terms of the
spin-Lorentzian ${\cal{A}}$, while
variation with respect to $e^{I}_{a}$ yields the usual vacuum Einstein
equations (Baez and Muniain, 1994). In our finsheaf of qausets scheme we
have not explicitly given finitary, causal and quantal gravitational
variables corresponding to the {\it
vierbein}. We only elaborated on ${\cal{A}}_{n}$ on $\vec{\cal{P}}_{n}$. It
is implicit however that, since the dynamically variable qauset-stalks of
$\vec{\cal{P}}_{n}$
locally define co-vector spaces of the form $\amg^{1}_{n}$, cross-sections of their
duals correspond to reticular, causal and quantal versions of the {\it vierbein}
observables\footnote{$4$-dimensionality and $\com$-coefficients arguments
aside, as we explained earlier.}. Hence, as we also made it clear in the introduction,
it is perhaps more accurate to
interpret our gauged finsheaf of qausets theoretic scheme $\vec{\cal{P}}_{n}$
as a finitary, causal and quantal replacement of the model
${\cal{P}}$ of curved spacetime structure in which classical Lorentzian gravity is
formulated as a gauge theory, rather than directly as a finitary, causal and
quantal substitute of GR {\it per se}. For instance, we have not even given the
finitary, causal and quantal analogues of the Einstein equations in
$\vec{\cal{P}}_{n}$. We do this in a forthcoming paper (Raptis,
2000{\it f}). Thus, it is perhaps more appropriate to view
$\vec{\cal{P}}_{n}$ as a kinematical structure that supports
the (germ of the) dynamics which is represented by the non-trivial ${\cal{D}}_{n}$
on it. To lay down the kinematics before the dynamics for qausets may be viewed as the first essential step in the development of the theory. For
example, Sorkin (1995), commenting on the influence that Taketani's
writings (1971) have exerted on the construction of causet theory and its
application to quantum gravity, stresses that ``{\it ...there is nothing
wrong with taking a long time to understand a structure
`kinematically' before you have a real handle on its dynamics}''. In
our case, there should be no confusion whatsoever when
one interprets $\vec{\cal{P}}_{n}$ directly as some kind of
`finitary and causal Lorentzian quantum
gravity' provided one remembers the remarks above.

($f^{'}$) Finally, Selesnick (1991), working on a correspondence
principle for Finkelstein's quantum causal net in (1988), speculated
on a
topos theoretic formalization of the net, where relativity and
the quantum will harmoniously coexist, and one that is more
fundamental than ${\bf Sh}(X)$-the category of sheaves of sets over
the topological spacetime
manifold $X$ in which both classical and
quantum field theories are currently  formulated. Indeed, a (quantum
?) topos organization of finsheaves of qausets may prove to be the
structure Selesnick anticipated. Searches for such a fundamental
`quantum topos' in which `quantum gravity' may be formulated rather
naturally have already been conducted from a slightly different point
of view than qauset theory proper (Raptis, 1998, 2000{\it c})\footnote{In a nutshell, it has been shown that the classical
({\it ie}, Boolean) topos of quaternion algebras $\quat$ corresponds to classical Minkowski spacetime $\cem$ (Trifonov,
1995). In (Raptis, 1998, 2000{\it c}) it has been argued that it is
precisely the (global) associativity of the quaternion product that
it is responsible for the flatness of $\cem$. In section 3 we too
entertained the idea that the (global) flatness of causets or
their corresponding qausets, which determine classical and quantal
versions of $\cem$, is due to their transitivity or their
associativity, respectively. In (Raptis, 1998, 2000{\it c}) a
non-classical ({\it ie}, non-Boolean or quantum ?) topos of
non-associative algebras is proposed to model a finitary curved
quantum causal space similar to the gauged finsheaves of qausets
that we proposed in the present paper or their respective finschemes
and their quantum topos organization in (Raptis, 2000{\it e}). The
latter may be regarded as a straightforward attempt at
arriving at `the true quantum topos for quantum gravity (and quantum
logic)' via qausets and their non-commutative schematic localizations.}. In any case, whatever this
elusive `quantum topos' structure turns out to be, it will certainly
shed more light on a very important mathematical question that has occupied
mathematicians
for some time now and which can be cast as the
following puzzling analogy: $\frac{\rm locales}{\rm
quantales}=\frac{\rm topoi}{\rm ?}$\footnote{Jim Lambek and Steve
Selesnick in private correspondence. Briefly, a topos or a locale
like, say, ${\bf Sh}(X)$, may be thought of as `a generalized pointless
topological space' and it may be interpreted as a universe of
variable sets (Selesnick, 1991, Mac Lane and
Moerdijk, 1992). Such an
interpretation is very welcome from the physical point of view that
we have adopted in the present paper according to which we are
looking for a mathematical structure to model the dynamical
variations of qausets in a way that does not commit itself to
spacetime as an inert background geometrical point-event
manifold $X$. However, much more work has to be done to put
finsheaves of qausets on a firm topos theoretic basis and study the
resulting structure's quantal features. As we said, significant progress in this
direction is analytically presented in (Raptis, 2000{\it
e}).}.

\section*{\normalsize\bf ACKNOWLEDGMENTS}

\indent The
second author (IR) acknowledges numerous exchanges on quantum
causality with
Roman Zapatrin (Torino), as well as a crucial
discussion about sheaf and scheme theoretic ring and algebra
localizations with Freddy Van Oystaeyen (Antwerp). He also wishes to
thank the Mathematics
Department of the University of Pretoria for a
postdoctoral research
fellowship with the help of which the present
work was written.

\section*{\normalsize\bf REFERENCES}

\sloppy{\noindent Alexandrov, A. D. (1956). {\it Helvetica Physica
Acta
(Supplement)}, {\bf 4}, 44.

\noindent Alexandrov,
A. D. (1967). {\it Canadian Journal of Mathematics},
{\bf 19}, 1119.

\noindent Alexandrov, P. S. (1956). {\it Combinatorial Topology},
vol. 1,  Greylock, Rochester, New York.

\noindent Alexandrov, P. S. (1961). {\it Elementary Concepts of
Topology}, Dover Publications, New York.

\noindent Ashtekar, A. (1986). {\it
Physical Review Letters}, {\bf 57}, 2244.

\noindent Baez,
J. C. and Muniain, J. P. (1994). {\it Gauge Fields, Knots
and
Quantum Gravity}, World Scientific, Singapore.

\noindent Bergmann,
P. G. (1957). {\it Physical Review}, {\bf 107}, 624.

\noindent Bombelli, L. and Meyer, D. (1989). {\it Physics Letters}, {\bf A141}, 226.

\noindent Bredon, G. E. (1967). {\it Sheaf Theory},
McGraw-Hill, New York.

\noindent Breslav, R. B., Parfionov,
G. N. and Zapatrin, R. R. (1999).
{\it Hadronic Journal}, {\bf 22},
225.

\noindent Bridgman, P. W. (1936). {\it The Nature of Physical
Theory},
Princeton University Press, Princeton.

\noindent
Candelas, P. and Sciama, D. W. (1983). {\it Physical Review D},
{\bf
27}, 1715.

\noindent Capozziello, S., Lambiase, G. and Scarpetta,
G. (2000). {\it
International Journal of Theoretical Physics}, {\bf
39}, 15.

\noindent Dimakis, A. and M\"uller-Hoissen,
F. (1999). {\it Journal of
Mathematical Physics}, {\bf 40},
1518.

\noindent Dimakis, A., M\"uller-Hoissen, F. and Vanderseypen,
F. (1995). {\it
Journal of Mathematical Physics}, {\bf 36},
3771.

\noindent Dodson, C. T. J. (1988). {\it Categories, Bundles
and Spacetime
Topology}, Kluwer Academic Publishers, Dordrecht.

\noindent Donoghue, J. F., Holstein, B. R. and Robinett,
R. W. (1984). {\it
Physical Review D}, {\bf 30}, 2561.

\noindent Donoghue, J. F., Holstein, B. R. and Robinett,
R. W. (1985). {\it General Relativity and Gravitation}, {\bf 17}, 207.

\noindent Einstein, A. (1924). {\it \"Uber den \"Ather},
{\it Schweizerische
Naturforschende Gesellschaft Verhanflungen},
{\bf 105}, 85-93
(English translation by Simon Saunders:
`{\it On the Ether}' in {\it
The Philosophy of Vacuum}, pp. 13-20, Eds. Saunders, S. and
Brown,
H., Clarendon Press, Oxford, 1991).

\noindent Einstein,
A. (1936). {\it Physics and Reality}, {\it J. Franklin
Institute},
{\bf 221}, 313.

\noindent Einstein, A. (1956). {\it The Meaning of Relativity},
Princeton
University Press.

\noindent Finkelstein, D. (1969). {\it Physical Review}, {\bf 184}, 1261.

\noindent Finkelstein, D. (1988). {\it International Journal of
Theoretical
Physics}, {\bf 27}, 473.

\noindent Finkelstein,
D. (1989). {\it International Journal of Theoretical
Physics}, {\bf
28}, 441.

\noindent Finkelstein, D. (1991). {\it Theory of Vacuum} in
{\it The Philosophy of Vacuum}, Eds. Saunders, S. and Brown, H.,
Clarendon Press, Oxford.

\noindent Finkelstein, D. (1996). {\it Quantum Relativity: a Synthesis of the Ideas of Einstein and Heisenberg},
Springer-Verlag, Berlin-Heidelberg-New York.

\noindent G\"ockeler, M. and Sch\"ucker, T. (1990). {\it Differential
Geometry, Gauge Theories and Gravity}, Cambridge University Press, Cambridge.

\noindent Hartshorne, R. (1983). {\it Algebraic Geometry}, 3rd
edition, Springer-Verlag, New York-Heidelberg-Berlin.

\noindent Mac Lane, S. and Moerdijk, I. (1992). {\it Sheaves in
Geometry and Logic: A First Introduction to Topos Theory}, Springer
Verlag, New York.

\noindent Mallios, A. (1998). {\it Geometry of Vector Sheaves: An
Axiomatic Approach to Differential Geometry}, vols. 1-2, Kluwer
Academic Publishers, Dordrecht (volume 3, which includes further physical
applications to Yang-Mills theories and gravity, is currently in the
process of publication).

\noindent Manin, Y. I. (1988). {\it Gauge Field Theory and
Complex
Geometry}, Springer-Verlag, New York.

\noindent Penrose, R. (1987). {\it Newton, Quantum Theory and Reality},
in {\it 300 Years of Gravitation}, Eds. Hawking, S. W. and Israel W., Cambridge
University Press, Cambridge.

\noindent Raptis, I. (1998). {\it Axiomatic Quantum Timespace Structure:
A Preamble to the Quantum Topos Conception of the Vacuum}, Ph.D. Thesis, University of Newcastle upon Tyne-UK.

\noindent Raptis, I. (2000{\it a}). {\it International Journal of
Theoretical Physics}, {\bf 39}, 1699.

\noindent Raptis, I. (2000{\it b}). {\it International Journal of Theoretical
Physics}, {\bf 39}, 1233.

\noindent Raptis, I. (2000{\it c}). {\it Non-Classical Linear Xenomorph as Quantum Causal Space and the Quantum Topos of Finitary Spacetime Schemes of Quantum Causal Sets}, in preparation (paper originally submitted to the {\it International Journal of Theoretical
Physics} (MS 990727.1), it is currently under revision and further development).

\noindent Raptis, I. (2000{\it d}). {\it Quantum Spacetime as a
Quantum Causal Set: What Happened to Operationalism ?}, in preparation.

\noindent Raptis, I. (2000{\it e}). {\it Non-Commutative Topology for
Curved Quantum Causality}, submitted to the {\it International Journal
of Theoretical Physics}.

\noindent Raptis, I. (2000{\it f}). {\it Locally Finite, Causal and Quantal Einstein Gravity}, in preparation.

\noindent Raptis, I. and Zapatrin, R. R. (2000). {\it International
Journal of Theoretical Physics}, {\bf 39}, 1.

\noindent Regge, T. (1961). {\it Nuovo Cimento}, {\bf 19}, 558.

\noindent Rideout, D. P. and Sorkin, R. D. (2000). {\it Physical
Review D}, {\bf 61}, 024002; e-print: gr-qc/9904062.

\noindent Robb, A. A. (1914). {\it Geometry of Space and Time}, (1st
edition), Cambridge University Press, Cambridge (2nd edition, CUP, 1936).

\noindent Rota, G-C. (1968). {\it Zeitschrift f\"ur
Wahrscheinlichkeitstheorie\/}, {\bf 2}, 340.

\noindent Selesnick, S. A. (1991). {\it International Journal of Theoretical
Physics}, {\bf 30}, 1273.

\noindent Selesnick, S. A. (1994). {\it Journal of Mathematical Physics},
{\bf 35}, 3936.

\noindent Selesnick, S. A. (1998). {\it Quanta, Logic and Spacetime:
Variations on Finkelstein's Quantum Relativity}, World Scientific,
Singapore.

\noindent Shafarevich, I. R. (1994). {\it Basic Algebraic Geometry 2}, 2nd
edition, Springer-Verlag, Berlin-Heidelberg-New York.

\noindent Sorkin, R. D. (1990). {\it Spacetime and Causal Sets}, in the
Proceedings of the SILARG VII Conference, Cocoyoc, Mexico (pre-print).

\noindent Sorkin, R. D. (1991). {\it International Journal of Theoretical
Physics}, {\bf 30}, 923.

\noindent Sorkin, R. D. (1995). {\it A Specimen of Theory Construction from
Quantum Gravity}, in {\it The Creation
of Ideas in Physics}, Ed. Leplin, J., Kluwer Academic Publishers, Dordrecht.

\noindent Stanley, R. P. (1986). {\it Enumerative Combinatorics}, Wadsworth
and Brook, Monterey, California.

\noindent Taketani, M. (1971). {\it Progress in
Theoretical Physics}, {\bf 50}, (Supplementum).

\noindent Torretti, R. (1981). {\it Relativity and Geometry}, Dover
Publications, New York.

\noindent Trifonov, V. (1995). {\it Europhysics Letters}, {\bf 32}, 621.

\noindent Van Oystaeyen, F. (2000{\it a}). {\it Algebraic Geometry for
Associative Algebras}, Marcel Dekker, New York.

\noindent Van Oystaeyen, F. (2000{\it b}). {\it Is the Topology of
Nature
Noncommutative ?} and {\it A Grothendieck-type of Scheme
Theory for Schematic
Algebras}, research seminars given at the
Mathematics Department of the University of Pretoria on 2-3/2/2000.

\noindent Van Oystaeyen, F. and Verschoren, A. (1981). {\it
Non-Commutative
Algebraic Geometry}, Lecture Notes in Mathematics
vol. 887, Springer-Verlag, Berlin.

\noindent Von Westenholz, C. (1981). {\it Differential Forms in
Mathematical
Physics}, Part 5, North Holland Publishing Company,
Amsterdam, New York,
Oxford.

\noindent Wheeler, J. A. (1964). In {\it Relativity, Groups and Topology},
Eds. De Witt, C. and De Witt, B. S., Gordon and Breach, London.

\noindent Zapatrin, R. R. (1998). {\it International Journal of Theoretical Physics}, {\bf 37}, 799.

\noindent Zapatrin, R. R. (2000). {\it Incidence Algebras of
Simplicial
Complexes}, submitted to {\it Pure Mathematics and its
Applications}, to appear; e-print math.CO/0001065.

\noindent Zeeman, E. C. (1964). {\it Journal of Mathematical Physics},
{\bf 5}, 490.

\noindent Zeeman, E. C. (1967). {\it Topology}, {\bf 6}, 161.

\end{document}